\documentclass[pre,aps,twocolumn]{revtex4}

\usepackage{amsmath}
\usepackage{amssymb}
\usepackage{graphicx}
\usepackage{subfigure}
\usepackage{algorithm}
\usepackage{algpseudocode}
\usepackage{bm}
\usepackage{color}
\usepackage{xcolor}

\usepackage{commath}
\usepackage{dcolumn}

\usepackage{hyperref}
\newcommand{\ceil}[1]{\lceil #1\rceil}

\newcommand{\avg}[1]{\langle #1\rangle}
\newcommand{\cut}[1]{{}}

\begin{document}

\title{Entropy Inflection and Invisible Low-Energy States: Defensive Alliance Example}

\author{
  Yi-Zhi Xu$^{1,4}$,
  Chi Ho Yeung$^{2}$,
  Hai-Jun Zhou$^{1,4,5}$,
  and David Saad$^{3}$
}

\affiliation{
  $^1$CAS Key Laboratory for Theoretical Physics, Institute of Theoretical Physics, Chinese Academy of Sciences, Beijing 100190, China \\
  $^2$Department of Science and Environmental Studies, The Education University of Hong Kong, Hong Kong\\
  $^3$Nonlinearity and Complexity Research Group, Aston University, Birmingham B4 7ET, United Kingdom\\
  $^4$School of Physical Sciences, University of Chinese Academy of Sciences, Beijing 100049, China \\
  $^5$Synergetic Innovation Center for Quantum Effects and Applications, Hunan Normal University, Changsha 410081, China
}
\date{\today}

\begin{abstract}
  Lower temperature leads to a higher probability of visiting low-energy states. This intuitive belief underlies most physics-inspired strategies for  addressing hard optimization problems. For instance, the popular simulated annealing (SA) dynamics is expected to approach a ground state if the temperature is lowered appropriately. Here we demonstrate that this belief is not always justified. Specifically, we employ the cavity method to analyze the minimum strong defensive alliance problem and discover a bifurcation  in the solution space, induced by an inflection point in the entropy--energy profile. While easily accessible configurations are associated with the lower-free-energy branch, the low-energy configurations are associated with the higher-free-energy branch within the same temperature range. There is a discontinuous phase transition between the high-energy configurations and the ground states, which generally cannot be followed by SA. We introduce an energy-clamping strategy to obtain superior solutions by following the higher-free-energy branch, overcoming the limitations of SA.
\end{abstract}

\maketitle

\begin{figure}[b]
\centering
\includegraphics[width=0.7\linewidth]{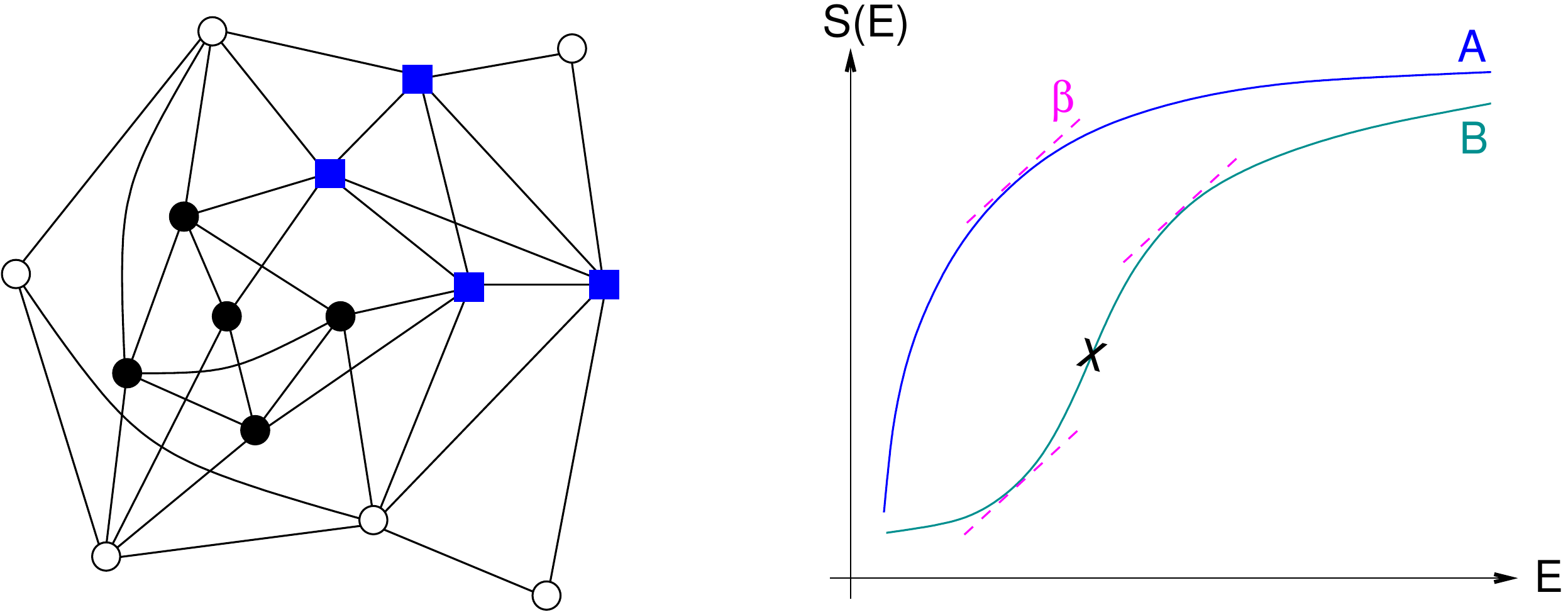}
\caption{
\label{fig:entropyphys}
(left) Two strong defensive alliance solutions for a small graph: the one denoted by filled circles has energy $E\!=\!5$; the other denoted by filled squares is the minimum alliance, $E\!=\!4$. (right) Two qualitatively different entropy curves $S(E)$: curve A is concave, its slope $\beta(E)$ decreases with energy $E$; curve B is non-concave, it has an inflection point (`X') at which the slope $\beta$ attains the maximum value.}
\end{figure}

Statistical physics associates the probability of visiting low-energy states with low temperatures. This has inspired the introduction of Metroplis-like algorithms~\cite{Metropolis:1953:ESC}, such as simulated annealing (SA), which sample low-energy configurations while gradually decreasing the temperature $T$, to progress towards equilibrium configurations close to the ground states~\cite{Kirkpatrick83optimizationby}. An implicit fundamental  assumption in SA is that the configuration entropy $S(E)$ is a concave function of the energy $E$ so that higher inverse temperature $\beta$ ($\equiv 1/T$) corresponds to lower $E$. In this work we show that for an important class of discrete-state systems, the entropy function is not always concave but is characterized by an inflection point that separates the concave higher-energy branch from the convex lower-energy branch (Fig.~\ref{fig:entropyphys}). Because low-energy configurations are associated with high microcanonical temperatures, they cannot be accessed by lowering the ambient temperature in a quasi-equilibrium manner.
Advanced multicanonical methods~\cite{Berg-Neuhaus-1991,Geyer-1991,Lyubartsev-etal-1992,Marinari-Parisi-1992,Hukushima-Nemoto-1996} that allow for an exchange between different temperatures will fail as well, being  rooted in the Boltzmann-Gibbs equilibrium framework, while the inflection of entropy means there must be a discontinuous phase transition between the ground states and high-energy configurations.

The exemplar optimization task adopted here is the minimum Strong Defensive Alliance (SDA) problem~\cite{Kristiansen-etal-2004}, a special case of finding substructures in a large graph~\cite{Jerrum-1992,balakrishnan2006discovering,Montanari-2015}. More specifically, one aims to identify the smallest group $A$ of vertices (the alliance) in the graph such that at least one half of the nearest neighbors of each alliance-vertex also belong to the alliance (Fig.~\ref{fig:entropyphys}). It is a nondeterministic polynomial hard (NP-hard) problem and has raised considerable interest among mathematicians~\cite{cami2006complexity, jamieson2009algorithmic,Yero-RodriguezVelazquez-2013}.
In statistical physics the SDA is closely related to the concepts of self-sustained clusters~\cite{Yeung-Saad-2013c,Rocchi2017a,Rocchi2017b} and metastable states~\cite{Lefevre-Dean-2001,Pagnani-etal-2003}, which are important for understanding the slow dynamics in spin systems. The synergetic excitation of a SDA may also drive rare but catastrophic cascading processes in real-world complex networks~\cite{Watts-2002}. In this paper we apply the cavity method of spin glasses~\cite{Mezard-etal-1986,Mezard-Parisi-2001,Mezard2009information} to the SDA problem. We find that the entropy function $S(E)$ is non-concave for relatively sparse graphs but recovers concavity when the graph becomes sufficiently dense. In addition, we develop a principled energy-clamping algorithm to construct nearly optimal alliance solutions. The insights gained in this study are applicable to a range of similar problems concerning densely connected subgraphs.

\emph{Strong Defensive Alliance}.-- Given a graph $G$ of $N$ vertices and $M$ edges, a non-empty subset $A$ of vertices is regarded as an alliance if and only if  at least half of the nearest neighbors of every vertex $i\in A$ are also in $A$. The minimum SDA problem aims to construct such an alliance of smallest cardinality, which requires a careful choice of vertices because SDA is a collective property of all vertices involved. For regular graphs in which every vertex has the same number $K$ of attached edges, the minimum alliance number is $2$ if $K\!=\!1, 2$ and it is equal to the graph's girth (the length of shortest loops) if $K\!=\!3, 4$. But for all $K\!\geq\!5$ the minimum SDA problem is intrinsically hard to solve, and the minimum alliance number is unknown and is difficult to bound~\cite{AraujoPardo-Barriere-2008}.  Here we apply methods and algorithms of statistical physics to tackle this challenging problem. For clarity we focus on regular random (RR) graphs, in which every vertex is linked to $K$ randomly drawn vertices. The formulation is generic and can be applied to other degree profiles. 

We cast the problem into a Hamiltonian form $E(\bm{c})\!=\! \sum_{i=1}^{N} \delta_{c_i}^1$, where $c_i \!=\! 1$ (the occupied state) if vertex $i$ belongs to the alliance and $c_i\! =\!0$ otherwise, and $\bm{c}\!\equiv\!(c_1, c_2, \ldots, c_N)$ denotes an occupation configuration of the $N$ vertices; the Kronecker symbol $\delta_{c}^{ c^\prime}\!=\!1$ if $c\!=\!c^\prime$ and $0$ otherwise. Let us denote by $\partial i$ the set of nearest neighbors of vertex $i$ and by $d_i\!\equiv\!|\partial i|$ its degree ($d_i\!=\!K$ if $G$ is regular). Each vertex $i$ gives rise to a constraint on $\bm{c}$: if $c_i=1$ then $\sum_{j\in \partial i} c_j \geq d_i/2$ must hold. Under these vertex constraints the partition function is
\begin{equation}
  \label{eq:Zbeta}
  Z(\beta)=\sum\limits_{\bm{c} \neq \bm{0}} \prod\limits_{i=1}^{N}
  \biggl[ \delta_{c_i}^0+e^{-\beta} \delta_{c_i}^1 
    \Theta\Bigl (\sum_{j\in \partial i} c_j - \frac{d_i}{2}  \Bigr)\biggr] \; ,
\end{equation}
where the Heaviside function $\Theta(x)\!=\!1$ if $x\!\geq\!0$ and $0$ otherwise. The all-zero crystalline state $\bm{0}\! \equiv  \! (0, 0, \ldots, 0)$ has been excluded from the summation since it does not correspond to an alliance. Each satisfying configuration (alliance) $\bm{c}$ contributes a term $e^{-n\beta}$ to $Z(\beta)$, where $n \equiv \sum_{i=1}^{N} c_i$ is the size of the alliance.

\begin{figure}
\centering
\subfigure[]{
\includegraphics[angle=270,width=0.475\linewidth]{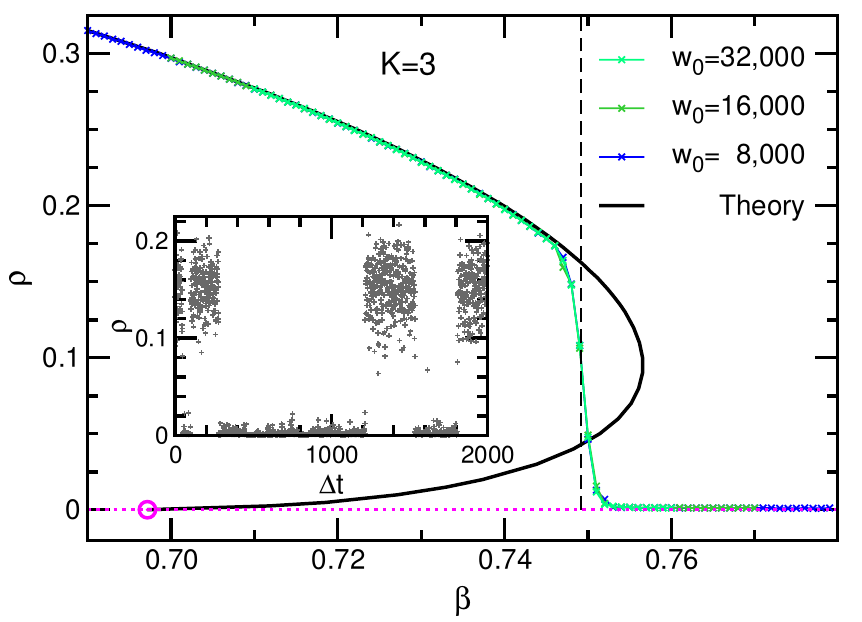}}
\subfigure[]{
\includegraphics[angle=270,width=0.475\linewidth]{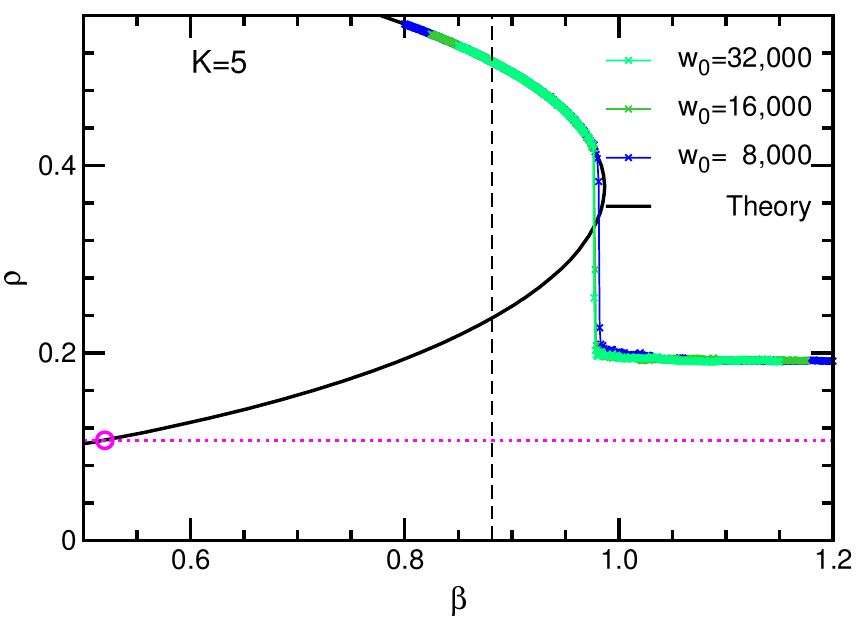}}
\caption{
\label{fig:TheoryMCMC}
The energy density $\rho$ of SDA identified by simulated annealing on a single RR graph of size $N\!=\!10^4$ and degree $K\!=\!3$ (a) or $K\!=\!5$ (b), as a function of ambient inverse-temperature $\beta$. Evolution trajectories obtained at three different waiting times $w_0$ are shown. 
The solid lines represent the theoretical curves of $\rho(\beta)$; the dotted horizontal lines and the circles mark the theoretical value of minimum energy density and the corresponding theoretical $\beta$. SA can reach the predicted minimum SDA size for $K\!=\!3$ but not for $K\!=\!5$. Dashed vertical lines mark $\beta$ values at the predicted discontinuous phase transition. The inset shows the fluctuation of $\rho$ at $\beta\!=\!0.75$ for $K\!=\!3$, with $\Delta t$ being the elapsed simulation time starting from an initial equilibrium configuration.
}
\end{figure}

\emph{Simulated annealing}.-- We implement a Markov-chain Monte Carlo dynamics to explore the SDA configuration space, which includes both single-vertex flipping and the simultaneous flipping of a connected chain or tree of vertices (details in~\cite{SM}). The Monte Carlo simulation runs for $w_0$ time steps at each ambient inverse temperature $\beta$ (one step contains $N$ flipping trials selected by importance sampling which guarantees detailed balance~\cite{Metropolis:1953:ESC,Newman-Barkema-1999}), and then $\beta$ is increased by a constant value $\varepsilon$ (e.g., $\varepsilon=0.001$). We run SA to identify SDA on two large RR graphs with degrees $K\!=\!3$ and $K\!=\!5$, and the results are shown in Fig.~\ref{fig:TheoryMCMC}(a) and \ref{fig:TheoryMCMC}(b) respectively. In both cases, the average SDA relative size $\rho$ (i.e. the energy density) first decreases gradually with increasing $\beta$ as anticipated; but it then violently fluctuates between two distinct levels as illustrated in the inset of Fig.~\ref{fig:TheoryMCMC}(a) when $\beta$ reaches a certain value $\beta_{SA}$ ($\approx\!0.75$ for $K\!=\!3$ and $\approx\!0.98$ for $K\!=\!5$); finally it settles at a low level as $\beta$ further increases. These simulation trajectories indicate the existence of a discontinuous phase transition, which is surprising since we do not expect the low-energy and minimum SDA solutions to be qualitatively different from the higher-energy SDA solutions.

For RR graphs with $K\!=\!3$ and $4$, the minimum SDA are triangular loops, which are frequently visited by the SA dynamics after $\rho$ drops to $\rho\!\sim\!1/N$. Since SA also saturates at a low energy level for the instance of $K\!=\!5$ (Fig.~\ref{fig:TheoryMCMC}b), one would naively claim the observed final value $\rho\!\approx\!0.187$ to be the minimum energy density. However, it turns out that the true minimum energy density is much lower ($\approx\!0.1067$).
Similar SA failures to visit low-energy configurations are observed on other graph instances~\cite{SM}. This might look unsurprising initially, since SA is well known to get trapped in metastable states if the low-energy configuration space fragments to an exponential number of disconnected ergodic domains~\cite{Montanari-RicciTersenghi-2004,Krzakala-Kurchan-2007}. However, our analysis does not support the emergence of such an explosive ergodicity-breaking phase transition at a high level of energy density~\cite{rivoire2004glass,Mezard-Montanari-2006,Krzakala-etal-PNAS-2007} (additional discussions in~\cite{SM}). Instead, we realize that the peculiar sudden drop followed by jamming as experienced by SA is due to another important but rarely discussed reason: the entropy curve as function of $\rho$ has an inflection point.

\emph{Mean field theory}.--  Random sparse graphs are characterized by long loops that diverge with graph size $N$. This allows us to consider the neighborhood of single vertices $i$ as tree-like, and for the neighboring vertices $j \!\in\! \partial i$ as mutually independent in the absence of $i$. Under this Bethe-Peierls factorization approximation~\cite{Mezard-etal-1986,Mezard-Parisi-2001,Mezard2009information}, the marginal probability $q_i$ of vertex $i$ belonging to the alliance is
\begin{equation}
  \label{eq:qi}
  q_i\!=\!\frac{e^{-\beta} \sum\limits_{\bm{c}_{\partial i}}
    \Theta\bigl(\sum\limits_{j\in\partial i} c_j -\frac{d_i}{2} \bigr)
    \prod\limits_{j\in\partial i} q_{j\rightarrow i}^{c_j,1} }
  {e^{-\beta} \sum\limits_{\bm{c}_{\partial i}} 
    \Theta\bigl(\sum\limits_{j\in\partial i} c_j\!-\!\frac{d_i}{2}\bigr)
    \prod\limits_{j\in\partial i}
    q_{j\rightarrow i}^{c_j,1}\!+\!\prod\limits_{j\in \partial i}
    (q_{j\rightarrow  i}^{0,0}\!+\!q_{j\rightarrow i}^{1,0})} \; .
\end{equation}
Here $\bm{c}_{\partial i} \!\equiv\!\{c_j \!:\! j\!\in\! \partial i\}$ denotes an occupation pattern of vertices in $\partial i$; and $q_{j\rightarrow i}^{c_j, c_i}$ is the probability of two nearest neighbors $i$ and $j$ being in states $c_i$ and $c_j$ simultaneously after lifting the constraint of vertex $i$. Following the same factorization approximation we obtain a closed set of self-consistent equations for the cavity probabilities $q_{j\rightarrow i}^{c_j, c_i}$:
\begin{eqnarray}
  q_{j\rightarrow i}^{0,0}  & \equiv &  q_{j\rightarrow i}^{0,1} = 
    \frac{1}{z_{j\rightarrow i}} \prod_{k\in \partial j\backslash i} 
    ( q_{k\rightarrow j}^{0,0}+q_{k\rightarrow  j}^{1,0}) \; ,
    \nonumber \\
    q_{j\rightarrow i}^{1,0} & = & \frac{e^{-\beta}}{z_{j\rightarrow i}} 
    \sum_{\bm{c}_{\partial j\backslash i}}
    \Theta\bigl(\sum\limits_{k\in\partial j\backslash i} c_k 
    -\frac{d_j}{2}\bigr) \prod\limits_{k\in \partial j\backslash i}
    q_{k\rightarrow j}^{c_k,1} \; ,
    \label{eq:BP}  \\
    q_{j\rightarrow i}^{1,1} & = & \frac{e^{-\beta}}{z_{j\rightarrow i}} 
    \sum\limits_{\bm{c}_{\partial j\backslash i}} 
    \Theta\bigl(\sum\limits_{k\in\partial j\backslash i} c_k + 1 
    -\frac{d_j}{2} \bigr) \prod\limits_{k\in \partial j\backslash i} 
    q_{k\rightarrow j}^{c_k,1} \; , \nonumber
\end{eqnarray}
where the set $\partial j\backslash i$ contains all the nearest neighbors of vertex $j$ except for $i$ and $\bm{c}_{\partial j\backslash i} \!\equiv\! \{c_k \!:\!  k \!\in\! \partial j\backslash i\}$;  $z_{j\!\rightarrow \!i}$ is the normalization constant ensuring that $\sum_{c_i,c_j} q_{j\rightarrow i}^{c_j, c_i} \! =\! 1$. This set of equations is collectively referred to as the belief-propagation (BP) equations~\cite{Mezard2009information}. 

Under the Bethe-Peierls approximation the expression for the free energy, $F\!\equiv\!-(1/\beta)\ln Z(\beta)$, of the system is~\cite{Mezard-Parisi-2001,Mezard2009information}
\begin{equation}
  \label{eq:free}
  F=\sum\limits_{i=1}^{N} f_{i+\partial i} - \sum\limits_{(i, j)\in G}f_{i j} \; ,
\end{equation}
where $f_{i+\partial i}$ is the contribution of vertex $i$ and all its attached edges, and $f_{i j}$ is the contribution of a single edge $(i, j)$. Because each edge $(i, j)$ contributes to both $f_{i+\partial i}$ and $f_{j+\partial j}$ its effect is subtracted once in Eq.~(\ref{eq:free}). The explicit expressions for $f_{i+\partial i}$ and $f_{i j}$ are:
\begin{eqnarray}
  f_{i+\partial i} & =& -\frac{1}{\beta}\ln\Bigl[e^{-\beta} 
    \sum\limits_{\bm{c}_{\partial i}}
    \Theta\bigl(\sum\limits_{j\in\partial i} c_j\!-\!\frac{d_i}{2}\bigr)
    \prod\limits_{j\in\partial i} q_{j\rightarrow i}^{c_j,1} \nonumber \\
    & & \quad \quad \quad \quad \quad + \prod\limits_{j\in \partial i}
    (q_{j\rightarrow  i}^{0,0}\!+\!q_{j\rightarrow i}^{1,0})\Bigr] \; , 
  \label{eq:fiandpi}
  \\
  f_{i j} &=& -\frac{1}{\beta}
  \ln\Bigl[\sum\limits_{c_i, c_j} q_{i\rightarrow j}^{c_i,c_j} 
    q_{j\rightarrow i}^{c_j,c_i} 
    \Bigr] \; .
  \label{eq:fij}
\end{eqnarray}

\begin{figure}
\centering
\subfigure[]{
\includegraphics[angle=270,width=0.475\linewidth]{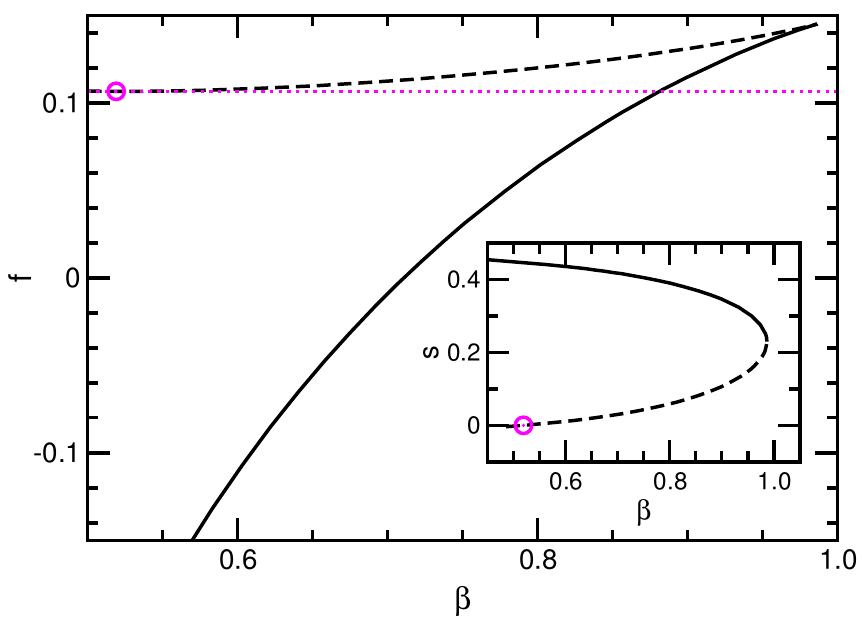}}
\subfigure[]{
\includegraphics[angle=270,width=0.475\linewidth]{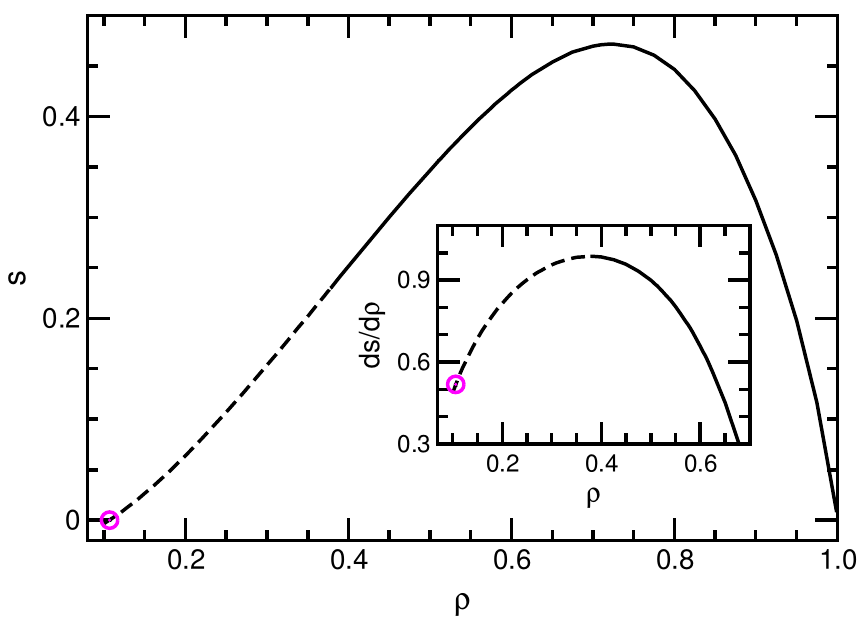}
}
\caption{
    \label{fig:K5sfrho}
The BP equation (\ref{eq:BP}) has two branches of fixed-point solutions for RR graphs of degree $K\!=\!5$. The lower-free-energy and higher-free-energy branches are drawn as solid and dashed lines, respectively. (a) Free energy density $f$ and entropy density $s$ versus inverse temperature $\beta$. (b) Entropy density $s$ and its first derivative ${\rm d}s/{\rm d} \rho$ versus energy density $\rho$.
The dotted horizontal line and the circles mark the ground-state free energy density and the corresponding $\beta$ and $\rho$ values.
}
\end{figure}

Equations (\ref{eq:qi})--(\ref{eq:fij}) constitute the replica-symmetric (RS) cavity theory~\cite{Mezard2009information} for the SDA problem. For the RR graph ensembles they can be further simplified after considering the vertex uniformity~\cite{SM}. We can iterate the BP equation either at fixed inverse temperature $\beta$, or at fixed energy density $\rho\!\equiv\!\sum_{i=1}^{N} q_i/N$ while adjusting $\beta$~\cite{SM}. The free energy density $f\!\equiv\!F/N$ and the entropy density $s\!\equiv\!(\rho-f)\beta$ are then computed at a fixed point of BP. 

\emph{Entropy Inflection}.-- The results of $f$, $s$ and $\rho$ for graphs with $K\!=\!5$ are shown in Fig.~\ref{fig:K5sfrho}, which are representative of all observed RR graphs with $3\le K \le 22$.
There is no fixed-point solution in the range of $\beta\!\geq\!0.9866$ ($K\!=\!5$) (see another explicit example for $K=3$ in~\cite{SM}); on the other hand there are two branches of BP fixed points when $\beta$ is smaller, a lower-free-energy (LFE) branch where $f$ increases while both $s$ and $\rho$  decrease with $\beta$, and a higher-free-energy (HFE) branch with opposite behaviors. Both branches are locally stable for fixed $\rho$ (microcanonical ensemble) but unstable with respect to message perturbations at fixed $\beta$ (canonical ensemble)~\cite{SM}; but because the HFE branch has a higher free energy it cannot be the dominant equilibrium state at a given ambient temperature (the canonical ensemble), even though its energy density $\rho$ is lower. This is a consequence of the much higher entropy of the LFE branch, arguably due to the large number of possible subset selections in larger alliances. The entropy density function $s(\rho)$ is monotonically increasing from zero to the maximum; it is initially convex until an inflection point is reached at $\rho_x\!=\!0.3775$ with a maximum slope $\beta_x\!=\!0.9866$. The entropy density approaches zero at $\rho_o\!=\!0.1067$, indicating that a minimum alliance contains only $0.1067 N$ vertices~\cite{kabashima1999statistical}.  The free energy density of the LFE branch exceeds that of the ground state at $\beta_{c}\!=\!0.8815$, implying a discontinuous equilibrium phase transition between the high-energy solutions ($\rho\!\approx\!0.511$) and the ground state ($\rho\!=\!\rho_o$) at this critical value $\beta_c$ (the corresponding value is $\beta_c\!=\!0.7491$ for $K\!=\!3$). The predicted discontinuous transition is exactly followed by SA on the $K\!=\!3$ graph instance ($\beta_{SA}=\beta_{c}$, Fig.~\ref{fig:TheoryMCMC}a) but it is much delayed by SA on the $K\!=\!5$ graph instance ($\beta_{SA} > \beta_c$, Fig.~\ref{fig:TheoryMCMC}b). Our numerical analysis~\cite{SM} reveals that the energetic and entropic barriers at the phase transition are finite and low for $K\!=\!3, 4$ but they are very high for $K\!=\!5$. The different SA behaviors of Fig.~\ref{fig:TheoryMCMC} are consistent with the fact that the minimum SDA problem is easy for $K\!=\!3, 4$ but NP-hard for $K\!\geq\!5$.

Qualitatively the same theoretical results are obtained for other RR graphs of degree $K\!\leq\!22$~\cite{SM}. The existence of an inflection point indicates nonequivalence of the canonical and the microcanonical statistical ensembles~\cite{Touchette-2015,Campa-etal-2009}.
Since the slope of $s(\rho)$ defines the intrinsic (microcanonical) inverse temperature, as the temperature $T$ decreases below $1/\beta_{x}$ the system is no longer capable of finding a matching stable equilibrium and will stay out-of-equilibrium if it has not fortuitously reached a ground state.
Notice that entropy-inflection is qualitatively different from the temperature-inflection phenomenon of~\cite{Schnabel-etal-2011} (see also \cite{Qi-Bachmann-2018}) as the latter does not result in a non-concave entropy curve. Non-concave microcanonical entropy was also discussed earlier in the contexts of ferromagnetic metastable states~\cite{Lefevre-Dean-2001,Pagnani-etal-2003} and constraint satisfiability problems~\cite{Zhou-Wang-2010}. Interestingly, we find that the entropy density $s(\rho)$ is concave for the entire physical region of $s\!\geq\!0$ (i.e., $\rho\!\geq\!\rho_o$) in high-$K$ graphs ($K\!\geq\!23$, see \cite{SM}). In these cases SA indeed successfully finds near-minimum SDA solutions~\cite{SM}.

\emph{Message-passing algorithm}.-- 
Because of entropy inflection, all configurations of low energy densities $\rho\!\in\!(\rho_o, \rho_x)$ are invisible in the Boltzmann-Gibbs equilibrium framework where temperature is gradually decreased. It appears that this discontinuity in the equilibrium energy spectrum causes extensive energetic and entropic barriers to the SA dynamics and prohibits the equilibrium transition from the high-energy configurations to the ground states (except the special $K\!=\!3,4$ cases for which the barriers are finite~\cite{SM}).
The optimization goal therefore is difficult to accomplish by quasi-equilibrium temperature annealing. One must adopt out-of-equilibrium search strategies. Inspired by the success of mean field theory in exploring the low-energy configuration space we propose a heuristic algorithm termed Clamp-Alliance (CA) for the SDA problem. This algorithm builds on the experiences of earlier message-passing methods~\cite{Mezard-etal-2002,Montanari-etal-ARXIV-2007,WongSaad08,Sulc-Zdeborova-2010,Xu-Zhou-2017} to perform BP-guided decimation with the objective size of the alliance set $A$ clamped at a low value $n_{obj}$. At each CA iteration: (1) the cavity probabilities $q_{i\rightarrow j}^{c_i, c_j}$ are updated several times, with a fine-tuned $\beta$ to ensure fixed mean energy $n_{obj}$; and (2) the occupation probability for every free vertex is evaluated by Eq.~(\ref{eq:qi}), and vertices $i$ with the lowest $q_i$ values are deemed  unsuitable for alliance membership and are fixed to be non-members ($c_i=0$). After the CA iteration stops an initial alliance set will be obtained. This set is then further refined until no other vertices can be removed. More details on the CA algorithm are provided in \cite{SM}. 

The performance of CA on some RR graphs is demonstrated in Table~\ref{tab:results}. By setting the objective (clamped) alliance size to $n_{obj}\!\approx\!\rho_o N$, we see that the solutions obtained by CA indeed have relative sizes $\rho$ close to the theoretically predicted minimum value $\rho_o$.
Let us point out that the CA algorithm can also be used to construct a near-minimum alliance set that is associated with a \emph{given} seed vertex. This latter problem might be particularly relevant for practical applications.

\begin{table}
  \caption{
    \label{tab:results}
    Mean energy density $\rho$ of alliances obtained by the Clamp-Alliance (CA) algorithm on $50$ RR graph instances of size $N\!=\!10^4$ and degree $K$, as compared with the theoretical minimum energy density $\rho_o$.
  }
  \centering
  \begin{tabular}{c|l|l|l|l|l|l}
    \hline\hline
    K & $5$ & $6$ & $7$ & $8$ & $9$ & $10$ \\ \hline
    CA & $0.126(3)$ & $0.062(3)$ & $0.237(2)$ & $0.159(2)$ & $0.295(1)$ & $0.227(2)$ \\
    $\rho_o$ & $0.1067$ & $0.0466$ & $0.2166$ & $0.1430$ & $0.2761$ & $0.2108$  \\
    \hline \hline
  \end{tabular}
\end{table}

\emph{Conclusion}.-- We studied a system with bifurcating branches of low and high free-energy configurations within the same temperature range, and revealed a discontinuous phase transition between the high-energy configurations and the non-crystalline ground states. Due to the presence of an inflection point in the entropy--energy profile of the system, the 
ground states are not associated with a low equilibrium temperature, and simulated annealing generally fails to follow the discontinuous phase transition to reach the ground state. Such a phenomenon is generic to the class of systems with an inflection point, which is crucial as we typically assume a monotonic and concave relation between energy and temperature, but do not verify the concavity property. We introduced an energy-clamping strategy to explore lowest-energy states located in the higher-free-energy branch, which overcomes the limitations of SA. This method can be extended to solve similar problems with a bifurcating configuration space. 
 
The conventional liquid--crystal phase transition is associated with a change in symmetry,
but the same does not hold for the present discontinuous phase transition between the high-energy configurations and the ground states, which originates from an inflection point of the entropy-energy profiles. It is interesting to search for such a distinct phase transition in finite-dimensional spin systems.

\begin{acknowledgments}
  YZX and CHY contributed equally to this work. Correspondence should be addressed to HJZ and DS. The following funding supports are acknowledged: Leverhulme Trust Grant RPG-2013-48 (DS); Research Grants Council of Hong Kong Grants 18304316 and 18301217) (CHY); National Natural Science Foundation of China Grants 11421063 and 11747601 (HJZ) and the Chinese Academy of Sciences Grant QYZDJ-SSW-SYS018) (HJZ). Numerical simulations were carried out at the HPC cluster of ITP-CAS and also at the Tianhe-2 platform of the National Supercomputer Center in Guangzhou. We thank Satoshi Takabe for valuable discussions.
\end{acknowledgments}


\clearpage

\appendix



\section{Simulated annealing (SA)}

Here we describe the details of the simulated annealing process. Without loss of generality we assume the input graph $G$ is connected. If instead $G$ is formed by two or even more connected components, each of these connected components can be treated separately. The SA process starts from an initial inverse temperature $\beta = \beta_{init}$, which is quite low (e.g., $\beta_{init}=10^{-3}$). The occupation configuration $\bm{c}=(c_1, c_2, \ldots, c_N)$ is initialized to be fully occupied, $c_i=1$ for all the vertices $i\in G$. Each occupied vertex contributes a unit energy, so the total energy of the initial configuration is $E(\bm{c})=N$. At each value of the inverse temperature $\beta$ the configuration $\bm{c}$ is allowed to evolve for a time $w_0$ through a sequence of single-vertex and multiple-vertex state flips, and the mean value of the configuration energies is recorded during this time window $w_0$. Then the inverse temperature is increased to $\beta \leftarrow  \beta + \varepsilon$ with $\varepsilon$ being a small value, e.g., $\varepsilon=0.001$ or $\varepsilon=0.01$. The SA process continues to run at this and later elevated $\beta$ values until the final value $\beta_{\text{final}}$ is reached, which is sufficiently high (e.g., $\beta_{\text{final}}=10$). The latest configuration $\bm{c}$ is then returned as the output of the SA evolution process. For the regular random (RR) graph instances studied in this work, we have checked that the subgraphs formed by the vertices in these final alliance solutions always have only a single connected component.

We adopt the Metropolis importance-sampling method to update the occupation configurations $\bm{c}$. In each elementary step of this Markov Chain Monte Carlo evolution dynamics: with probability $p_s$ a single-vertex state flip is attempted, and with the remaining probability $p_m = 1-p_s$ a multiple-vertex state flip is attempted; and then the evolution time $t$ advances by the incremental change $\delta t=1/N$ irrespective of whether the proposed change to $\bm{c}$ was accepted or rejected. One unit time of the SA evolution therefore corresponds to $N$ consecutive flipping trials. Let us emphasize that the SA process generates a stochastic trajectory within the space of strong defensive alliance (SDA) solutions; at any evolution time $t$ the vertex set formed by the occupied vertices of $\bm{c}$ is always a valid alliance. 

We set $p_s=p_m=0.5$ in all our SA simulations. The SA algorithm applicable to $K$-regular graphs is accessible from the webpage {\tt power.itp.ac.cn/\~{}zhouhj/codes.html}.

\subsection{Single-vertex state flip}

A single-vertex flipping trial consists of proposing a state change $c_i \rightarrow 1-c_i$ for a vertex $i$ of the graph, under the constraint that the initial configuration $\bm{c}$ and the updated configuration $\bm{c}^\prime$ are both valid alliances. For the initial configuration $\bm{c}$, let us denote the set of all the flippable vertices from $c_i=1$ to $c_i=0$ as $V_{1\rightarrow 0}$ and the set of all flippable vertices from $c_j=0$ to $c_j=1$ as $V_{0\rightarrow 1}$; similarly, for the updated configuration $\bm{c}^\prime$ the sets of flippable $(1\rightarrow 0)$ and $(0\rightarrow 1)$ vertices are denoted as $V_{1\rightarrow 0}^\prime$ and $V_{0\rightarrow 1}^\prime$, respectively. The cardinality of a vertex set (say $V$) is denoted as $|V|$. We conduct single-vertex flipping trials following the rule of importance sampling, which guarantees detailed balance:
\begin{enumerate}
\item[1.] Generate a uniform real random number $r$ in $[0, 1)$.
\item[2.] If $r < \frac{|V_{1\rightarrow 0}|}{|V_{1\rightarrow 0}|+e^{-\beta} |V_{0\rightarrow 1}|}$, randomly choose a flippable occupied vertex $i$ from set $V_{1\rightarrow 0}$ and propose a flip from $c_i=1$ to $c_i=0$; otherwise randomly choose a flippable empty vertex $j$ from $V_{0\rightarrow 1}$ and propose a flip from $c_j=0$ to $c_j=1$. 
\item[3.] Accept this single-vertex flip proposal and the associated new configuration $\bm{c}^\prime$ with probability $A_s(\bm{c}\rightarrow \bm{c}^\prime)$, whose precise expression being
  \begin{equation}
    \hspace*{0.8cm} A_s(\bm{c}\rightarrow \bm{c}^\prime) = \min\Bigl(1, \ \frac{|V_{1\rightarrow 0}| + e^{-\beta} |V_{0\rightarrow 1}|}{|V_{1\rightarrow 0}^\prime|+ e^{-\beta} |V_{0\rightarrow 1}^\prime|} \Bigr) \; ,
  \end{equation}
  otherwise keep the old configuration $\bm{c}$.
\end{enumerate}

\subsection{Multiple-vertex state flip for a regular graph of degree $K=3$}

\begin{figure}
  \centering
  \includegraphics[width=0.5\textwidth]{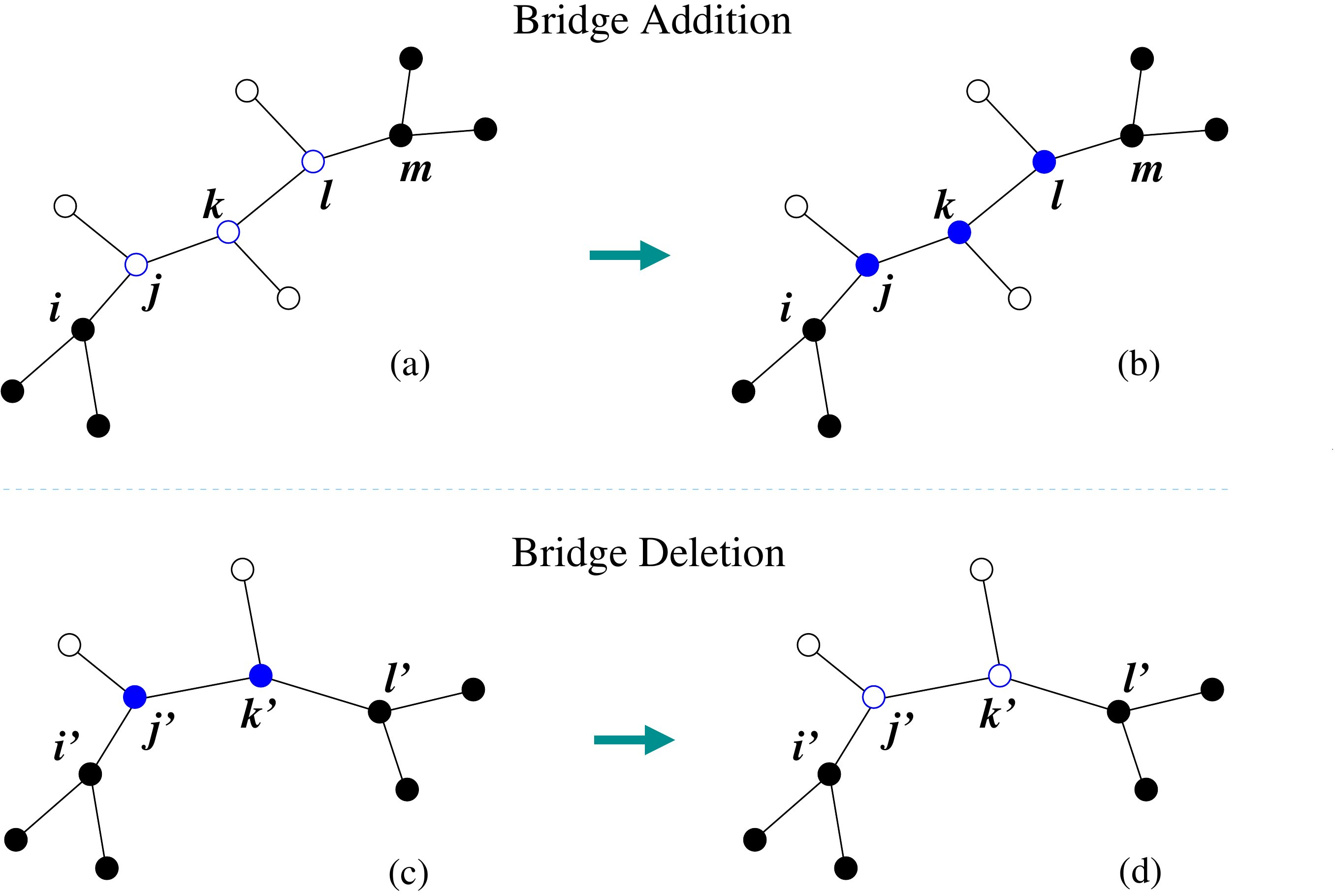}
  \caption{
    \label{fig:bridgeK3}
    Bridge addition and deletion processes for regular graphs of degree $K=3$. (a) Vertices $i$ and $m$ serve as the two anchors for an empty bridge $(j, k, l)$ of length $n_1=3$ of an occupation configuration $\bm{c}$. (b) After an empty bridge is flipped to be an occupied bridge, a new configuration $\bm{c}^\prime$ is formed, and vertices $i$ and $m$ then serve as the two anchors of an occupied bridge $(j, k, l)$ in this new configuration. (c) and (d): An occupied bridge $(j^\prime, k^\prime)$ of length $n_2=2$ anchored to vertices $i^\prime$ and $l^\prime$ is flipped to an empty bridge.
     }
\end{figure}

To better explain the adopted multiple-vertex flipping trials we first consider the special case of a regular graph of degree $K=3$ (i.e., a $3$-regular graph). We define the concepts of empty and occupied bridges as follows: An empty bridge for a $3$-regular graph is a path formed by $n\!\geq\!2$ different empty vertices $j_1, j_2, \ldots, j_n$ such that: (1) the whole path is connected to two and only two occupied vertices (called the bridge anchors, e.g., vertices $i$ and $m$ in Fig.~\ref{fig:bridgeK3}a) by exactly two edges, attached to the start and end vertices ($j_1$ and $j_n$), and there is no other neighboring occupied vertex to the whole path except the two bridge anchors, and (2) there is no other edge between any two vertices of this path except for the $n\!-\!1$ edges linking these $n$ empty vertices into a path. Similarly, an occupied bridge for a $3$-regular graph is a path formed by $m\!\geq\!2$ different occupied vertices $k_1, k_2,\ldots, k_m$ such that: (1) all these $m$ vertices have exactly two occupied nearest neighbors, (2) there is no other edge between any two vertices of this path except for the $m\!-\!1$ edges linking them into a path, and (3) the start and end vertices ($k_1$ and $k_m$) of the bridge are connected to two \emph{different} occupied vertices (the bridge anchors, e.g., vertices $i^\prime$ and $l^\prime$ of Fig.~\ref{fig:bridgeK3}c) with each of these two anchors having three occupied nearest neighbors.

For the initial configuration $\bm{c}$, let us denote by $B_{0\rightarrow 1}$ the set formed by all the start or end vertices of all the empty bridges, and by $B_{1\rightarrow 0}$ the set formed by all the start or end vertices of all the occupied bridges. Similarly, the two sets of bridge terminal vertices for the updated configuration $\bm{c}^\prime$ are denoted as $B_{0\rightarrow 1}^\prime$ and $B_{1\rightarrow 0}^\prime$, respectively. Notice that if a vertex $j\in B_{0\rightarrow 1}$ is flipped to the state $c_j=1$ all the other vertices of the associated empty bridge must be flipped to be occupied as well; similarly if a vertex $j^\prime \in B_{1\rightarrow 0}$ is flipped to the state $c_{j^\prime}=0$ all the other vertices of the associated occupied bridge must be flipped to empty as well (Fig.~\ref{fig:bridgeK3}).

To construct an empty bridge of configuration $\bm{c}$ we proceed as follows: (1) Draw an empty vertex (say $j$ of Fig.~\ref{fig:bridgeK3}a) from set $B_{0\rightarrow 1}$ and regard it as the start of an empty bridge. (2) Then randomly select an empty neighbor (say vertex $k$) of $j$ and add it to the bridge. (3) Determine whether to stop or to continue: if $k$ has two occupied neighbors, the bridge construction is regarded as a failure and is stopped; otherwise if $k$ has only one occupied neighbor, the bridge construction is regarded as successful and is stopped; otherwise $k$ has no occupied neighbor, then the bridge is extended by adding an randomly chosen empty neighbor (say vertex $l$) different from $j$ to the bridge and then step (3) is repeated. If the construction of the empty bridge is successful, the last added vertex (e.g., $l$ in Fg.~\ref{fig:bridgeK3}a) must be connected to a single occupied vertex.

The construction of an occupied bridge is slightly simpler: (1) Draw an occupied vertex (say $j^\prime$ of Fig.~\ref{fig:bridgeK3}c) from set $B_{1\rightarrow 0}$ and regard it as the start of an occupied bridge. (2) Then add a neighboring occupied vertex (say $k^\prime$) with exactly two occupied neighbors to the bridge. (3) Continue this bridge extension process if necessary, until an anchor vertex (say $l^\prime$ in Fig.~\ref{fig:bridgeK3}c) is reached. The constructed bridge is regarded as successful if the two anchor vertices $i^\prime$ and $l^\prime$ of the bridge are not identical.

We conduct the multiple-vertex flipping trial from $\bm{c}$ to $\bm{c}^\prime$ according to the following rule of importance sampling, which guarantees detailed balance:
\begin{enumerate}
\item[1.] Generate a uniform real random number $r$ in $[0, 1)$.
\item[2.] Perform bridge addition or bridge deletion:
  \begin{enumerate}
  \item[(2.1).] If $r<\frac{|B_{0\rightarrow 1}|}{|B_{0\rightarrow 1}|+|B_{1\rightarrow 0}|}$, then randomly choose an empty vertex $j$ from set $B_{0\rightarrow 1}$ and construct an empty path starting from $j$ following the above-mentioned protocol. If the constructed path is not a valid empty bridge, keep the old configuration $\bm{c}$. If this path is a valid empty bridge, then flip all the vertices in this bridge to be occupied and accept the updated configuration $\bm{c}^\prime$ with the following probability
    %
    \begin{eqnarray}
      & & \hspace*{0.8cm}A_{m}^{0\rightarrow 1}(\bm{c}\rightarrow \bm{c}^\prime) =
      \label{eq:Am01}
      \\
      & & \quad \quad \quad \min\biggl(1, \ \frac{|B_{0\rightarrow 1}|+|B_{1\rightarrow 0}|}{|B_{0\rightarrow 1}^\prime|+|B_{1\rightarrow 0}^\prime|} 2^{n_b-1} e^{-\beta n_b} \biggr) \; ,
      \nonumber 
    \end{eqnarray}
    %
    where $n_b$ denotes the length of the constructed bridge; otherwise keep the old configuration $\bm{c}$.
  \item[(2.2).] Otherwise $r\geq \frac{|B_{0\rightarrow 1}|}{|B_{0\rightarrow 1}|+|B_{1\rightarrow 0}|}$, then randomly choose an occupied vertex $j^\prime$ from set $B_{1\rightarrow 0}$ and extend an occupied path starting from $j^\prime$, following the above-mentioned protocol. If the constructed path is not a valid occupied bridge, keep the old configuration $\bm{c}$. If this path is a valid occupied bridge, then flip all the vertices in this bridge to be empty and accept the updated configuration $\bm{c}^\prime$ with the following probability
    %
    \begin{eqnarray}
      & & \hspace*{0.8cm} A_{m}^{1\rightarrow 0}(\bm{c}\rightarrow \bm{c}^\prime) = 
   \label{eq:Am10}
   \\
   & & \quad \quad \quad \min\biggl(1, \ \frac{|B_{0\rightarrow 1}|+|B_{1\rightarrow 0}|}{|B_{0\rightarrow 1}^\prime|+|B_{1\rightarrow 0}^\prime|} 2^{1- n_b} e^{\beta n_b} \biggr) \; ,
   \nonumber 
    \end{eqnarray}
    where $n_b$ again denotes the length of the constructed bridge; otherwise keep the old configuration $\bm{c}$.
  \end{enumerate}
\end{enumerate}

\subsection{Multiple-vertex state flip for a general graph}

The bridge addition and deletion processes can be extended to a general graph, but Eqs.~(\ref{eq:Am01}) and (\ref{eq:Am10}) have to be modified accordingly. Here we describe the extended bridge flipping processes from one configuration $\bm{c}$ to another configuration $\bm{c}^\prime$. For simplicity  we assume the input graph $G$ to be $K$-regular (i.e., each vertex having $K$ nearest neighbors). Let us denote $\theta\equiv \ceil{\frac{K}{2}}$. An empty vertex (say $i$) is regarded as a candidate start/end of a possible empty bridge if $i$ has exactly $\theta \!-\!1$ occupied neighbors. Notice that if such a vertex $i$ is flipped to the state $c_i\!=\!1$ one of its empty neighbors must also be flipped. The sets of such empty terminal vertices of the initial configuration $\bm{c}$ and of the updated configuration $\bm{c}^\prime$ are denoted as $B_{0\rightarrow 1}$ and $B_{0\rightarrow 1}^\prime$, respectively. An occupied vertex $j$ is regarded as a candidate start/end of a possible occupied bridge if (1) $j$ has exactly $\theta$ occupied neighbors and, (2) one of these occupied neighbors (say $k$) has exactly $\theta$ occupied neighbors itself while all the other occupied neighbors have more than $\theta$ occupied neighbors. Notice that if $j$ is flipped to $c_j\!=\!1$ the occupied neighbor $k$ must also be flipped. The sets of such occupied terminal vertices in $\bm{c}$ and $\bm{c}^\prime$ are denoted as $B_{1\rightarrow 0}$ and $B_{1\rightarrow 0}^\prime$, respectively.

To construct an empty bridge for the configuration $\bm{c}$ we proceed as follows: (1) Set index $l=1$ and draw an empty vertex $i_l$ from the set $B_{0\rightarrow 1}$ and consider it as the start of an empty bridge. (2) Construct a set $C_{i_l}$ for the newly added vertex $i_l$, which contains all the empty vertices $k$ satisfying the following properties: (a) $k$ is a nearest neighbor of $i_l$ but it is not a nearest neighbor of any other existing vertices of the bridge (to avoid loop formation), (b) $k$ has not yet been added to the bridge, and (c) $k$ has either $\theta\!-\!1$ or $\theta\!-\!2$ occupied neighbors. If set $C_{i_l}=\emptyset$, the bridge construction is regarded as a failure and it is terminated; otherwise randomly draw an empty vertex $i_{l+1}$ from $C_{i_l}$ and add it to the empty chain. (3) Set $l\leftarrow l+1$. If the last added vertex has exactly $\theta\!-\!1$ occupied neighbors, the bridge construction is regarded as successful and it is terminated, otherwise go back to step (2) to try to further elongate the empty bridge. If this bridge construction process is successfully finished, we obtain an empty bridge $(i_1, i_2, \ldots, i_{n_b})$ involving $n_b \geq 2$ empty vertices. Because of the randomness in extending this empty bridge, we assign it a ``surprising'' scale as
\begin{equation}
  \label{eq:Wsurprise}
  W^0_{i_1, i_2, \ldots, i_{n_b}}  = \prod\limits_{l=1}^{n_b-1} |C_{i_l}| \; ,
\end{equation}
where $|C_{i_l}|$ denotes the cardinality of vertex set $C_{i_l}$. Notice that the set $C_{i_l}$ for index $l\!\geq\!2$ is affected by the vertices $i_1, i_2, \ldots, i_{l-1}$ of the bridge.

To construct an occupied bridge for the configuration $\bm{c}$ is easier than constructing an empty bridge. Let us refer to an occupied vertex $j$ as being critical occupied if it has exactly $\theta$ occupied neighbors (so it has to be flipped to $c_j\!=\!0$ if any one of its occupied neighbors is flipped). Then a candidate occupied bridge is generated in the following way: (1) Set index $l=1$ and draw an occupied vertex $i_1$ from the set $B_{1\rightarrow 0}$ and consider $i_1$ as the start of an occupied bridge. (2) Add the only critically occupied nearest neighbor (say vertex $i_2$) of $i_1$ to the bridge and increase the index to $l=2$. (3) If the newly added vertex $i_{l}$ has only one critically occupied neighboring vertex (i.e., $i_{l-1}$) the candidate bridge is constructed and the process is terminated; if $i_l$ has more than two critically occupied neighboring vertices the bridge construction is regarded as failed and it is terminated; otherwise $i_{i}$ has exactly two critically occupied neighbors (one is $i_{l-1}$, the other one is denoted as $i_{l+1}$), then we add $i_{l+1}$ to the bridge, increase index $l\leftarrow l+1$, and repeat the last step (3) to further elongate the occupied bridge if necessary. After this bridge construction process is successfully finished, we obtain a candidate bridge $(i_1, i_2, \ldots, i_{n_b})$ involving $n_b\geq 2$ occupied vertices. To check whether this occupied chain is a valid bridge, we flip all the vertices of this chain to be empty. If every occupied nearest neighboring vertex of this chain still has $\theta$ or more occupied nearest neighbors itself after this chain has been flipped to empty, then the chain is regarded as a valid bridge and its ``surprising'' scale as an empty bridge is computed according to Eq.~(\ref{eq:Wsurprise}), otherwise it is regarded as a failure. After this check all the vertices in the chain is flipped back to be occupied.

Given an occupation configuration $\bm{c}$, if we decide to perform a multiple-vertex flipping trial (which occurs with probability $p_m$), then
\begin{enumerate}
\item[1.] With probability $\frac{|B_{0\rightarrow 1}|}{|B_{0\rightarrow 1}| + |B_{1\rightarrow 0}|}$ it is a bridge addition trial: an empty chain $(i_1, i_2, \ldots, i_{n_b})$ of variable length $n_b\geq 2$ is generated according to the above-mentioned protocol and, if it is a valid empty bridge, the whole bridge is flipped and accepted with probability
  %
  \begin{eqnarray}
     & & A_{m}^{0\rightarrow 1}(\bm{c}\rightarrow \bm{c}^\prime) =  
    \label{eq:Am01gen}  \\
    & & \quad \min\biggl(1, \ \frac{|B_{0\rightarrow 1}|+|B_{1\rightarrow 0}|}{|B_{0\rightarrow 1}^\prime|+|B_{1\rightarrow 0}^\prime|} W^0_{i_1, i_2, \ldots, i_{n_b}}  e^{-\beta n_b} \biggr) \; . \nonumber 
  \end{eqnarray}
\item[2.] With the remaining probability $\frac{|B_{1\rightarrow 0}|}{|B_{0\rightarrow 1}| + |B_{1\rightarrow 0}|}$ it is a bridge deletion trial: an occupied chain $(i_1, i_2, \ldots, i_{n_b})$ of variable length $n_b\geq 2$ is generated according to the above-mentioned protocol and, if it is a valid occupied bridge, the whole bridge is flipped and accepted with probability
%
  \begin{eqnarray}
    & & A_{m}^{1\rightarrow 0}(\bm{c}\rightarrow \bm{c}^\prime) = 
    \label{eq:Am10gen}
    \\
& & \quad \min\biggl(1, \ \frac{|B_{0\rightarrow 1}|+|B_{1\rightarrow 0}|}{|B_{0\rightarrow 1}^\prime|+|B_{1\rightarrow 0}^\prime|} \frac{1}{W^0_{i_1, i_2, \ldots, i_{n_b}}} e^{\beta n_b} \biggr) \; . \nonumber
\end{eqnarray}
    %
  Let us emphasize again that $W^0_{i_1, i_2, \ldots, i_{n_b}}$ in Eq.~(\ref{eq:Am10gen}) is the surprising scale of the resulting empty bridge $(i_1, i_2, \ldots, i_{n_b})$ \emph{after} the flip.
\end{enumerate}

\subsection{Extending bridge-flipping into tree-flipping}

\begin{figure*}
  \centering
  \subfigure[]{
    \includegraphics[angle=270,width=0.322\textwidth]{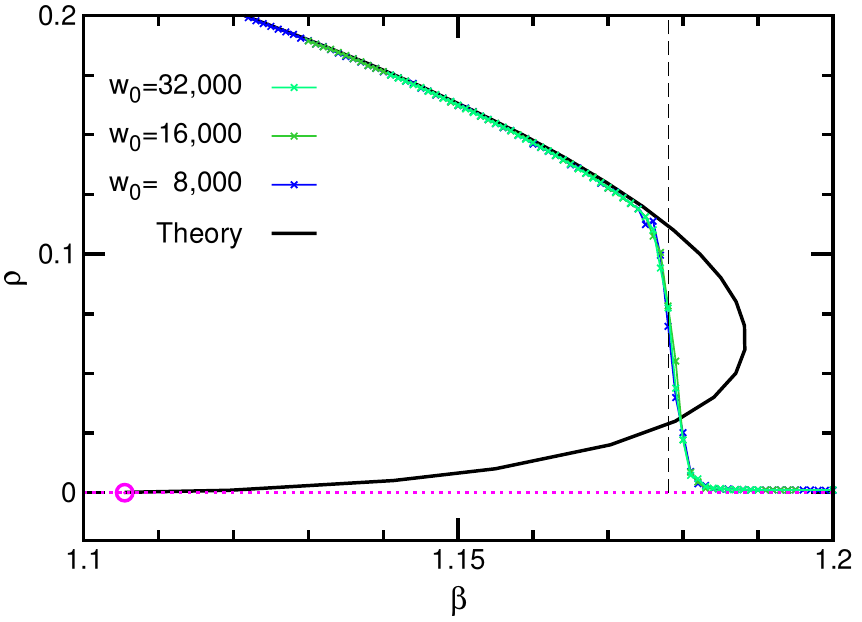}}
  \subfigure[]{
    \includegraphics[angle=270,width=0.322\textwidth]{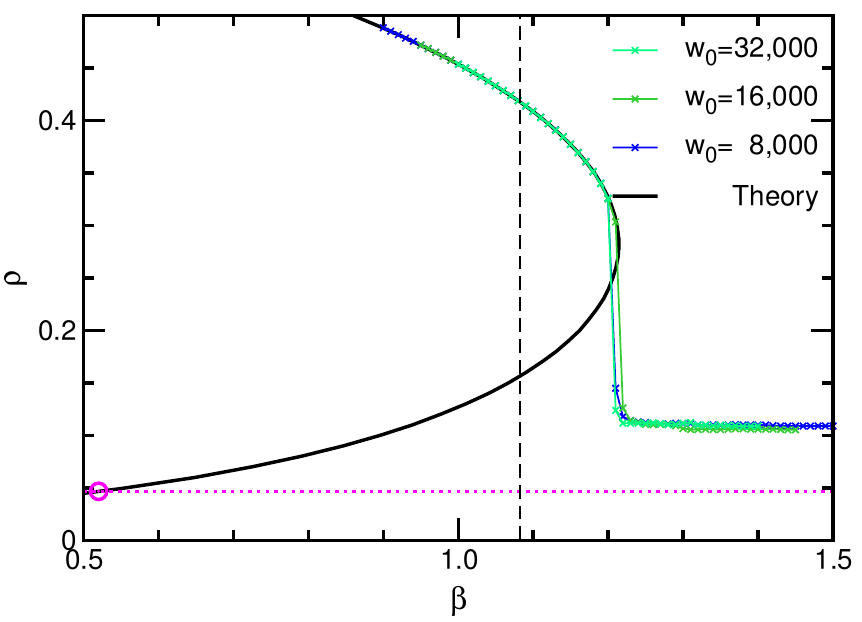}}
  \subfigure[]{
    \includegraphics[angle=270,width=0.322\textwidth]{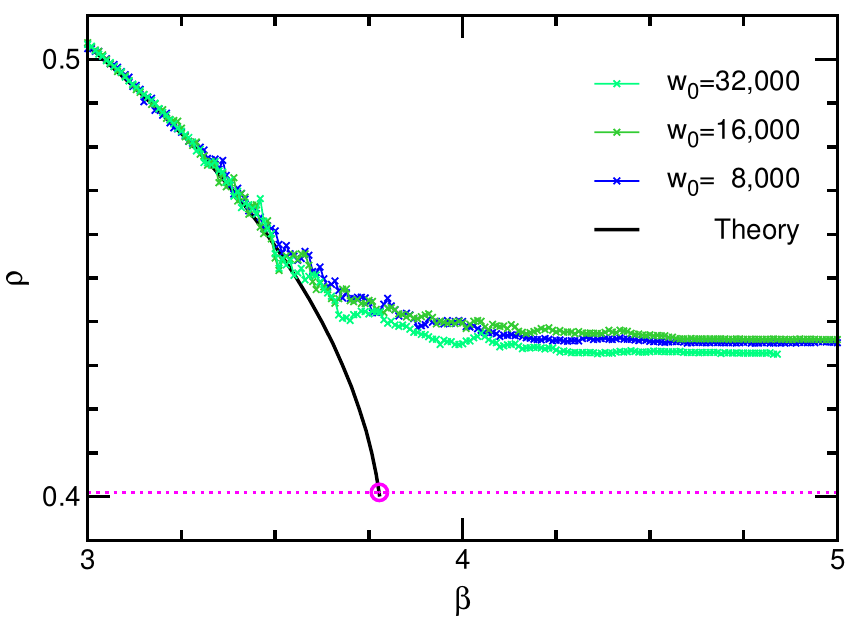}}
  \caption{
    The same as Fig.~2 of the main text. Simulated annealing results on a single RR graph of size $N=10^4$ and degrees $K=4$ (a), $K=6$ (b), and $K=25$ (c) are compared with theoretical predictions. Evolution trajectories obtained at three different waiting times $w_0$ are shown. The inverse temperature is denoted by $\beta$ while $\rho$ is the relative size of alliances. In each panel the bold solid line represents the theoretical curve of $\rho$ versus $\beta$, and the circle and dotted horizontal line mark the predicted ground-state energy density. An equilibrium discontinuous phase transition is predicted to occur for the RR ensembles of $K\leq 22$. The phase transition point, obtained from the corresponding free energy values, is marked by the dotted vertical line in (a) and (b).
  } 
  \label{fig:MCMCK4K6K25}
\end{figure*}

The bridge-flipping process of the preceding subsection can be extended into tree-flipping process with some modifications. We define a connected subgraph of the $K$-regular graph as a flippable occupied tree (FOT) if the following conditions are satisfied: (1) the FOT forms a connected subgraph without any internal loops; (2) each vertex $i$ of this FOT is occupied ($c_i=1$) and has exactly $\theta$ occupied neighbors; (3) flipping all the vertices in this FOT will not force any other vertices in the graph to be flipped. Similarly a flippable empty tree (FET) is defined as a connected subgraph without any internal loops with the following additional properties: (1) every vertex $i$ in this FET is empty ($c_i=0$); (2) every leaf vertex of this FET is connected to exactly $\theta\!-\!1$ occupied external vertices (which do not belong to the FET) and one vertex in the FET; (3) every non-leaf vertex $j$ of this FET is connected to $d_j \in \{0, 1, \ldots, \theta\}$ other vertices of the FET and exactly $\theta\!-\!d_j$ occupied external vertices.

According to the above definitions, a FOT can be flipped to be a FET without disturbing the states of all other vertices, and a FET can be flipped back to be a FOT without the need of flipping any additional empty vertices. We have implemented this tree-flipping process under the condition of detailed balance. It turns out that the resulting numerical code is much slower than that of the bridge-flipping process. When testing on the RR graph instances of degrees $K=3, 4, 5, 6$ we found that the tree-flipping SA algorithm produces quantitatively very similar results as the bridge-flipping SA algorithm. For example, in the case of $K\!=\!5$, the dramatic energy drop occurs at $\beta_{SA}\!\approx\!0.979$ and the final energy level is $\rho\!\approx\!0.180$; the corresponding values for the $K\!=\!6$ case are $\beta_{SA}\!\approx\!1.205$ and $\rho\!\approx\!0.102$. Because tree-flipping does not significantly improves the performance of SA, in this work we choose to use bridge-flipping as the multiple-vertex flipping mechanism. Detailed analysis of the tree-flipping SA algorithm will be reported in a follow-up paper.

\subsection{Discussions on the performance of SA}

SA dynamical results obtained for random $K$-regular graphs are shown in Fig.~2 and Fig.~\ref{fig:MCMCK4K6K25}. When each vertex has only $K=3$ (Fig.~2a) or $K=4$ (Fig.~\ref{fig:MCMCK4K6K25}a) nearest neighbors, the SA trajectory can successfully reach a minimum alliance solution, after experiencing an abrupt drop in energy density $\rho$, at a certain critical value of inverse temperature $\beta$ predicted by the cavity theory (marked by the vertical dashed line of Fig.~2a and Fig.~\ref{fig:MCMCK4K6K25}a). The simulated annealing behaviors observed on the $3$- and $4$-RR graphs indeed fully agree with the theoretical prediction. This algorithmic success can be well explained.

\begin{figure*}
  \centering
  \subfigure[]{
    \includegraphics[angle=270, width=0.322\textwidth]{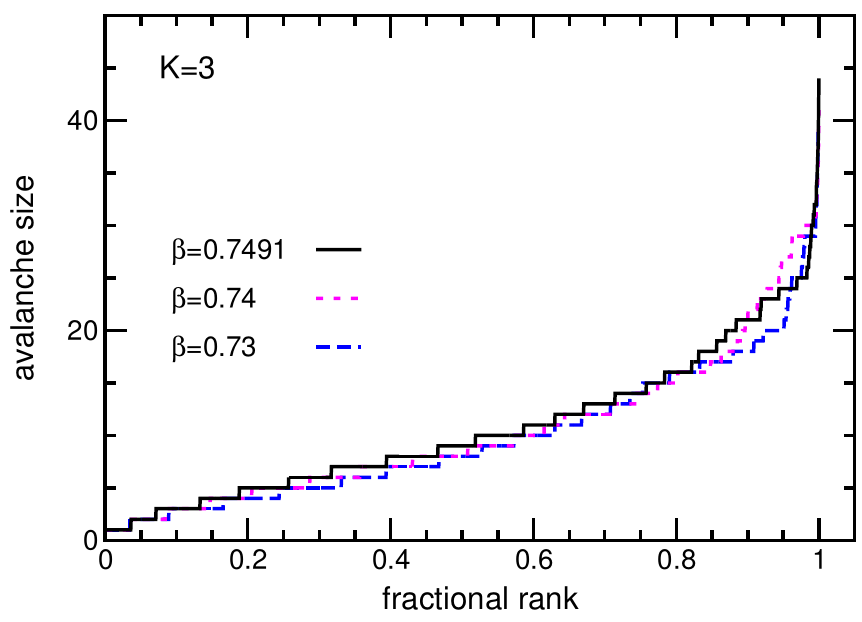}}
\hspace{0.2cm}
  \subfigure[]{
    \includegraphics[angle=270,width=0.322\textwidth]{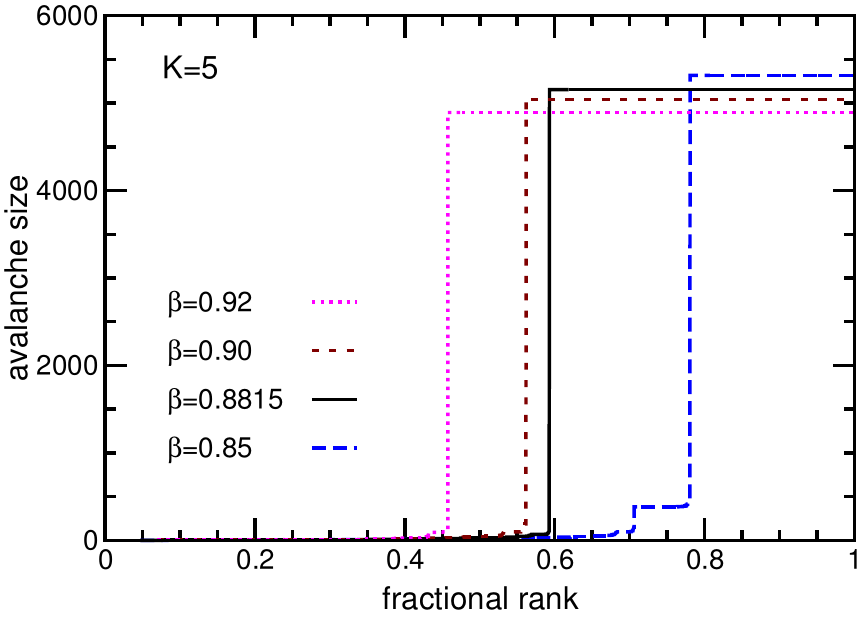}}
  \caption{
    The size of avalanches induced by flipping a single occupied vertex $i$ on an alliance configuration $\bm{c}$. The size $a_i$ of this avalanche is defined as the total number of flipped vertices (including $i$). The $n$ occupied vertices in $\bm{c}$ are ranked with index $r=1, 2, \ldots, n$ according to the avalanching effect and the relative rank is simply $r/n$. The results of avalanche size for (a) a $3$-RR graph instance of size $N=10^4$, at three different values of $\beta$ (the predicted phase transition point is $\beta_c=0.7491$), with the alliance size $n$ being $n=2129$ ($\beta=0.73$), $n=1930$ ($\beta=0.74$) and $n=2017$ ($\beta=0.7491$); and (b) a $5$-RR graph of size $N=10^4$, at four different $\beta$ values ($\beta_c=0.8815$), with alliance size being $n=5318$ ($\beta=0.85$), $n=5156$ ($\beta=0.8815$), $n=5043$ ($\beta=0.90$) and $n=4894$ ($\beta=0.92$). The big jumps in (b) correspond to all the vertices in the graph being in the empty state.}
  \label{fig:AvalancheK3K5}
\end{figure*}

Given an occupation configuration $\bm{c}$ we refer to a vertex $i$ as being critical if this vertex is occupied ($c_i\!=\!1$) and it has exactly $\theta =\ceil{\frac{K}{2}}$ occupied nearest neighbors. A critical vertex will collapse to the empty state if any one of its occupied nearest neighbors if flipped to be empty. Since $\theta\!=\!2$ in the cases of $3$- and $4$-RR graphs, a critical vertex $i$ has at most two critical nearest neighbors and so its flipping will immediately affect at most two other occupied vertices. If vertex $j$ is such a critical nearest neighbor of $i$, it will have at most one other critical nearest neighbor besides $i$, so the induced flipping of $j$ will immediately affect at most one additional occupied vertex, and the same applies for the critical neighbor of $j$ and so on.
By this analysis we see the critically occupied vertices of $\bm{c}$ form some simple paths (without self-loops) which do not share any vertex. The occupied bridges sampled by the SA algorithms are just some of these critical paths. If such an occupied chain is flipped as a whole, a new occupation configuration $\bm{c}^\prime$ of lower energy will be obtained. On the other hand, suppose there is an empty vertex $j$ which has only a single occupied nearest neighbor and we flip $j$ to be occupied (i.e., from $c_j=0$ to $c_j=1$). Then we only need to flip one of its empty nearest neighbors (say $k$) to make $j$ satisfy the constraint of being in the alliance. If vertex $k$ again only has one occupied nearest neighbor (which is $j$), then we only need to flip one of the empty nearest neighbors to stabilize $k$, and the same applies for the neighbors of $k$ and so on. After this chain extension process stops, a new occupation configuration $\bm{c}^\prime$ of increased energy is reached.

By repeatedly applying the above-mentioned chain flipping and single-vertex flipping processes, any occupation configuration of a $3$- or $4$-RR graph can be reached from any another occupation configuration, meaning that the algorithm can reach all configurations of the system and it is an ergodic algorithm. To guarantee detailed balance property of the SA dynamics, we have further restricted the flipped chain to be a bridge (there should be no internal loop among the vertices of this chain, and flipping of this chain should not cause any of the connected occupied vertices to be unstable) but these restrictions do not affect the ergodic property of the SA dynamics. In a random graph the typical length of a shortest-distance path between two vertices grows logarithmically with the graph size $N$. We therefore expect the energy gap of flipping an empty bridge to be at most of order $\log(n_b)$, with $n_b$ being the bridge length. In our simulations $n_b$ exceeded $20$ only very rarely. We have formulated a percolation theory (to be described in a following paper) to compute the mean value of $n_b$; this theory predicts that, for $3$- and $4$-RR graphs, $n_b$ is only of $O(1)$ even for an infinite graph ($N\rightarrow\infty$). The energy barrier of bridge flipping can therefore be easily overcome. This property together with the ergodicity property of SA for the $3$- and $4$-RR graphs explain why the SA evolution trajectories in Fig.~2a and Fig.~\ref{fig:MCMCK4K6K25}a  abruptly drop at the theoretical predicted phase transition point to visit a ground state.

Results for RR graphs of degrees $K=5$ and $K=6$, shown in Fig.~2b and Fig.~\ref{fig:MCMCK4K6K25}b, exhibit an abrupt drop of energy density $\rho$ during the SA evolution process; this does not occur at the predicted equilibrium phase transition point (the vertical dashed line of Fig.~2b and Fig.~\ref{fig:MCMCK4K6K25}b), but close to the predicted entropy inflection point. After this much delayed drop in energy the SA evolution trajectory still fails to reach the energy level of ground states but is trapped at a much higher energy level. It seems that the energy barriers are high in these graph instances and the SA evolution dynamics with only single-vertex and bridge (or tree) flips is unable to overcome these barriers, leading to effective ergodicity-breaking in the SA process.

To see why ergodicity in the configuration space of the $K$-RR graph ($K\geq 5$) might be severely broken at low energy levels, let us investigate the consequence of flipping an occupied vertex $i$ (from $c_i=1$ to $c_i=0$). If an occupied nearest neighboring vertex $j$ of $i$ is critical (that is, having exactly $\theta$ occupied nearest neighbors), $j$ will no longer be marginally stable and it will collapse to the empty state ($c_j=0$). Since $\theta\geq 3$ vertex $j$ may itself be connected to more than one critically occupied vertex besides $j$, and its collapse may then induce the collapse of two or more (up to $\theta-1$) critically occupied nearest neighbors, and so on. When this avalanche process finally stops and we count the remaining occupied vertices, with high probability the whole alliance solution has collapsed! This single-vertex flipping may therefore induce a complete collapsing behavior as demonstrated in Fig.~\ref{fig:AvalancheK3K5}b on a $5$-RR graph instance, and it is prohibited because the all-empty configuration does not correspond to a valid alliance solution. This global collapsing behavior is dramatically different from the situation observed on a $3$-RR graph (Fig.~\ref{fig:AvalancheK3K5}a), for which the avalanche size is always finite ($<50$) at any value of $\beta$. 

For the $5$--RR graph, as long as the inverse temperature $\beta$ exceeds $0.8$, we observed that a finite fraction of the occupied vertices in \emph{every} visited equilibrium configurations are completely blocked (flipping any one of these occupied vertices will cause the collapse of the whole alliance solution). We have developed a percolation theory to quantitatively understand this strong blocking phenomenon (to be reported in the follow-up paper). The equilibrium dynamics of the system is therefore severely restricted. For such a blocked vertex (say $i$) to be flippable, the system has to rearrange itself (through many local single-vertex or multiple-vertex flips) into a suitable configuration in which $i$ is no longer blocked; but with the relaxation of vertex $i$ some other vertices will be blocked and the evolution trajectory will still be strongly restricted. In other words, there is a high degree of dynamical heterogeneity among the vertices: some of the vertices can be easily flipped while the others are completely blocked, and every vertex changes between these two coarse-grained states over time. The entropic barrier associated with an extensive number of blocked vertices may make it impossible for the SA evolution process to realize the huge energy drop at the predicted discontinuous phase transition point $\beta_{c}$. Instead the SA dynamics enters into the ``super-cooled'' non-equilibrium region (see Fig.~2b and Fig.~\ref{fig:MCMCK4K6K25}b) as $\beta$ exceeds $\beta_c$.

The simulation results of Fig.~\ref{fig:MCMCK4K6K25} on a RR graph of degree $K=25$ demonstrate a smooth decrease of energy density $\rho$ with inverse temperature $\beta$, in agreement with the theoretical prediction of the absence of a discontinuous phase transition in RR graph ensembles of degree $K\geq 23$. However, at $\beta \approx 3.5$ the SA evolution trajectories start to deviate from the theoretical $\rho(\beta)$ curve, possibly due to the waiting times $w_0$ used in the SA dynamics becoming shorter than the characteristic system relaxation time. It may also be possible that the low-energy configurations (with $\rho<0.45$) of this $25$-RR graph instance are in the spin glass phase. This possibility deserves to be thoroughly explored in future investigations.

\section{Theoretical expressions for a regular-random (RR) graph}
\label{sec:RSuniform}

The BP equations~(3) of the main text can be solved iteratively (see the following section). For a RR graph of degree $K$, due to the uniformity of vertex properties; it turns out that the fixed-point cavity probability distributions on all edges are identical. Therefore the BP equations~(3) for the RR graph ensemble can be simplified to
\begin{subequations}
  \label{eq:BPsimple}
  \begin{align}
    q^{(0,0)}= & q^{(0,1)} =
    \frac{1}{z} (q^{(0,0)}+q^{(1,0)})^{K-1} \; , \\
    q^{(1,0)}= & \frac{e^{-\beta}}{z} 
    \sum\limits_{d\geq \frac{K}{2}}^{K-1} 
    C_{K-1}^d (q^{(1,1)})^d (q^{(0,1)})^{K-1-d} \; ,\\
    q^{(1,1)}= & \frac{e^{-\beta}}{z} 
    \sum\limits_{d\geq \frac{K}{2}-1}^{K-1}C_{K-1}^d
    (q^{(1,1)})^d (q^{(0,1)})^{K-1-d} \; ,
  \end{align}
\end{subequations}
where $C_{n}^{m} \equiv \frac{n!}{m! (n-m)!}$, and $z$ is the normalization constant. The corresponding marginal occupation probability (simply $\rho$) for a vertex is
\begin{equation}
  \label{eq_q1}
  \rho =
  \frac{e^{-\beta} \sum\limits_{d\geq \frac{K}{2}}^{K}
    C_{K}^d (q^{(1,1)})^d  (q^{(0,1)})^{K-d}}{(q^{(0,0)}+q^{(1,0)})^{K}
    + e^{-\beta}\sum\limits_{d\geq \frac{K}{2}}^{K}\! C_{K}^d
    (q^{(1,1)})^d (q^{(0,1)})^{K-d}} \; .
\end{equation}

Equations~(\ref{eq:BPsimple}a)-(\ref{eq:BPsimple}c) can be analytically solved for the simplest non-trivial case of degree $K=3$, and the solution demonstrates the existence of an inflection point in the entropy--energy profile. Let us first simplify the notation by introducing
\begin{equation}
  \label{eq:abcd}
  a = q^{1,1} \; , \quad 
  b = q^{0,1} \; , \quad 
  c = q^{1,0} \; , \quad 
  d = q^{0,0} \; .
\end{equation}
For the case of $K=3$ the BP equation (\ref{eq:BPsimple}) can be written as
\begin{equation}
  \label{eq:k3}
b  = d = \frac{1}{z} (c+d)^{2} \; , \quad 
a  = \frac{e^{-\beta}}{z} (a^2 + 2ab) \; , \quad 
c = \frac{e^{-\beta}}{z} a^2 \; .
\end{equation}
One can re-arrange Eqs.~(\ref{eq:BPsimple}a)-(\ref{eq:BPsimple}c) to obtain the exact solution of cavity probabilities, and subsequently the free energy and the entropy. In this case, by using Eq.~(\ref{eq:k3}), we obtain
\begin{eqnarray}
  \frac{a}{c} & = & 1 + 2\frac{b}{a}\; ,
  \label{eq_ac}  \\
  \frac{b}{a} & = &
  \frac{\bigl(\frac{b}{a}+\frac{c}{a}\bigr)^2}{e^{-\beta}
    \bigl(1+2\frac{b}{a}\bigr)}\; .
  \label{eq_ba}
\end{eqnarray}
Let us denote $x=a/b$. From the above equations we obtain the following equation for $x$
\begin{equation}
  \label{eq:quartic}
  (x^2+x+2)^2 - e^{-\beta}(x+2)^3 = 0 \; .
\end{equation}

\begin{figure}
  \centering
  \includegraphics[angle=270, width=0.4\textwidth]{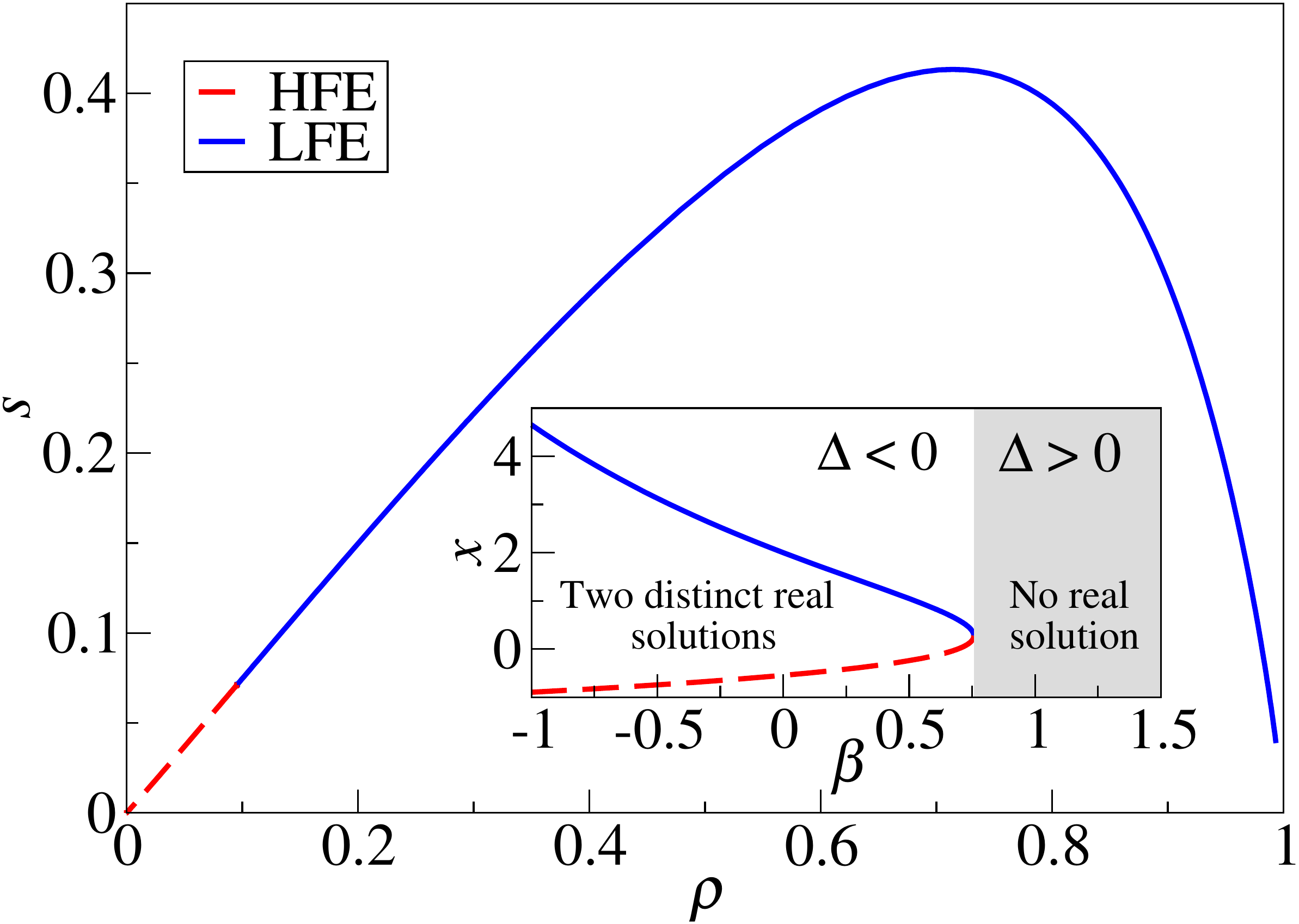}
  \caption{
    \label{fig:k3andk25}
    The non-concave entropy density function $s(\rho)$ for the RR graphs of degree $K=3$, obtained from Eqs.~(\ref{eq:rhok3}) and (\ref{eq:sk3}), with the solution of $x$ from Eq.~(\ref{eq:quartic}). Inset: the solution of $x$ given by Eq.~(\ref{eq:quartic}), which shows that there is a range of $\beta$ values with no real solution for $x$. Similar to Fig.~3 of the main text, the corresponding BP fixed-point solutions form higher (HFE) and lower-free-energy (LFE) branches.
 }
\end{figure}

The energy density $\rho$, the free energy density $f$, and the entropy density $s$ can also be expressed in terms of $x$ as
\begin{subequations}
  \begin{align}
    \rho &= \frac{e^{-\beta}(x^3+3x^2)}{\bigl(\frac{x^2}{x+2}+1\bigr)^3
      + e^{-\beta}(x^3+3x^2)} \; ,
    \label{eq:rhok3}
    \\
    f &= -\frac{1}{\beta}\ln\Bigl[\bigl(\frac{x^2}{x+2}+1\bigr)^3 
      + e^{-\beta}(x^3+3x^2)\Bigr] \nonumber \\
    & \quad \quad \quad 
    +\frac{3}{2\beta}\ln\Bigl[1+x^2+ \frac{2x^2}{x+2} \Bigr] \; ,    
    \label{eq:fk3}\\
    s &= \beta (\rho-f) \; .
    \label{eq:sk3}
  \end{align}
\end{subequations}
By solving the quartic equation in~(\ref{eq:quartic}) at a given value of $\beta$, one can obtain both lower and higher free-energy solutions, real and complex. Since only the real solutions are relevant in the present case, we first write the determinant $\Delta$ of Eq.~(\ref{eq:quartic}) as $\Delta = \Delta_1^2 - 4\Delta_0^3$, where $\Delta_0= -72 e^{-\beta} + 49$ and $\Delta_1 = 432 e^{-2\beta} + 1512 e^{-\beta} - 686$. One can then solve Eq.~(\ref{eq:quartic}) explicitly for $\beta$ when $\Delta=0$, which gives
\begin{equation}
  \beta = -\ln\Bigl(\frac{19\sqrt{57}-135}{18}\Bigr) \approx 0.757 \; .
\end{equation}
When the determinant of the quartic equation $\Delta>0$, i.e. $\beta>0.757$, there is no real solution for $x$ in Eq.~(\ref{eq:quartic}) and consequently for the cavity probabilities. On the the other hand, there are two distinct solutions when $\Delta<0$, i.e. $\beta<0.757$ (see Fig.~\ref{fig:k3andk25}). By using Eqs.~(\ref{eq:rhok3})--(\ref{eq:sk3}), we can plot the entropy--energy profile $s(\rho)$ with the two solutions of $x$ sharing the same temperature range $\beta<0.757$. The concave branch of $s(\rho)$ is obtained from one of the solutions, while the other leads to the convex branch. The inflection point of $s(\rho)$ locates at the value of $\rho$ for which $\beta= 0.757$.

\begin{figure}
  \centering
  \includegraphics[width=0.4\textwidth]{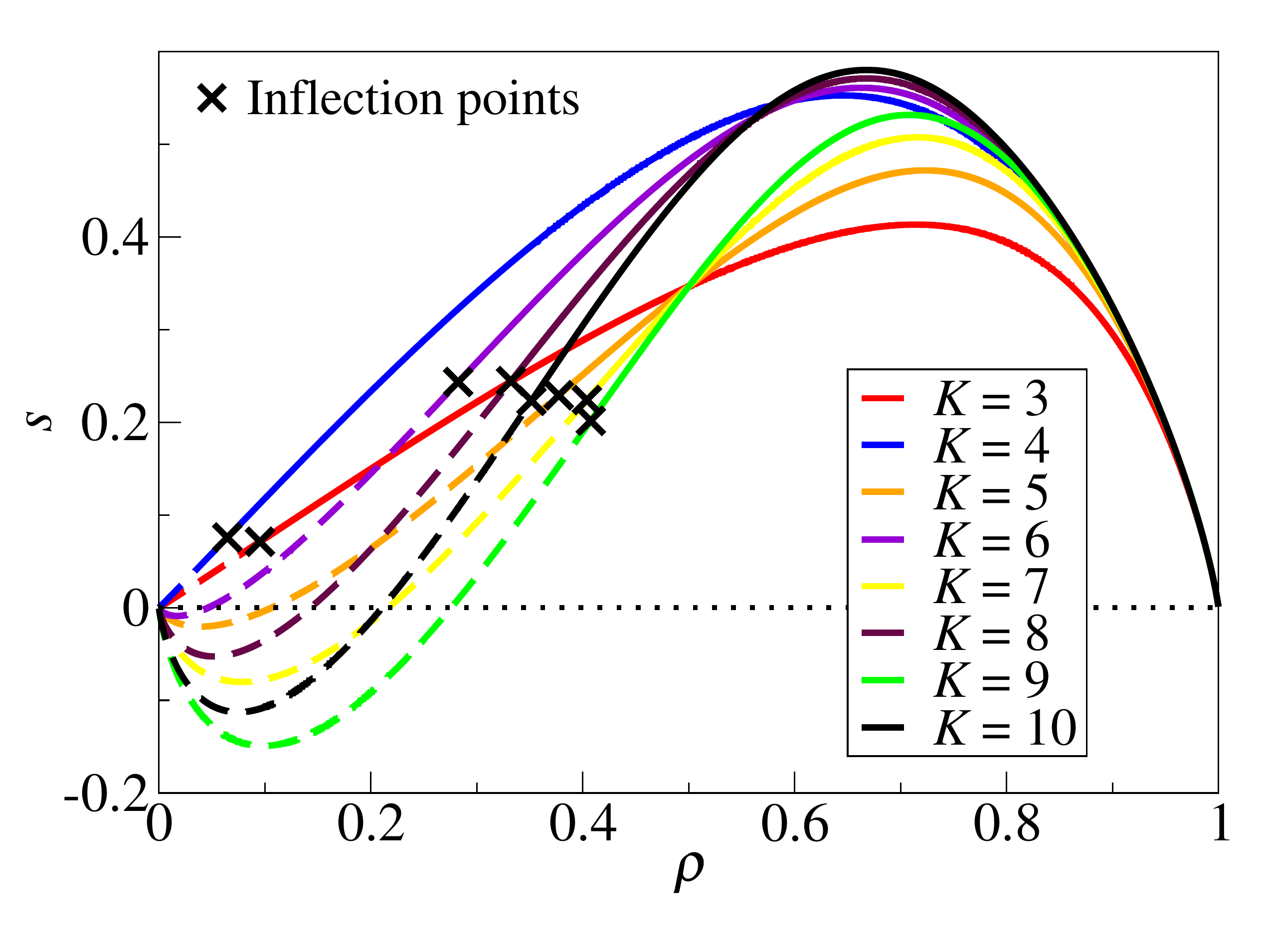}
  \caption{
    The analytical results of the entropy--energy profile $s(\rho)$ for RR graphs with $K=3$ to $K=10$, obtained by the cavity theory in Eq. (2)-(6) of the main text. The inflection points are indicated by the cross symbols $\times$, and the dashed lines correspond to the convex regime of $s(\rho)$.
  }
  \label{fig:entropyK3-10}
\end{figure}

As for related work, we notice that Ref.~\cite{Schnabel-etal-2011} discussed the inflection point of the inverse temperature, but not that of the entropy, and specifically conclude that the entropy function will be concave in the thermodynamic limit. For the SDA system studied here, the non-concavity of the entropy function persist in the thermodynamic limit.

In addition to the emergence of an inflection point ($\rho=\rho_x$), the entropy density at low $\rho$ values ($\rho<\rho_o$) becomes negative and unphysical for RR graphs with $K\ge 5$, as shown in Fig.~\ref{fig:entropyK3-10}. Since the number of configurations at a given energy density $\rho$ is of order $e^{N s(\rho)}$, a negative value of entropy density indicates that low-energy configurations of $\rho<\rho_o$ are non-existent in a typical RR graph instance.
Therefore, we define the SDA ground states to be the states of minimal SDA with non-negative entropy. According to this definition, since the entropy for the cases of $K=3,4$ is always positive (see Fig.~\ref{fig:entropyK3-10}), the SDA ground states are characterized by $\rho \gtrsim 0$. As discussed in the main text, the ground states for $K=3,4$ are states with occupied triangular loops; if we denote $n$ to be the number of alliance nodes, the ground states are characterized by $n=3$, such that $\rho_o=n/N \gtrsim 0$ in a system with large $N$. In this case, $s(\rho_o)$ is infinitesimally positive, implying that there may be more than one ground state, i.e. more than one state with a different occupied triangle loops, which is consistent with the results of the statistics of loops in RR graphs obtained in \cite{marinari2004circuits}.

\begin{figure}
  \centering
  \includegraphics[width=0.4\textwidth]{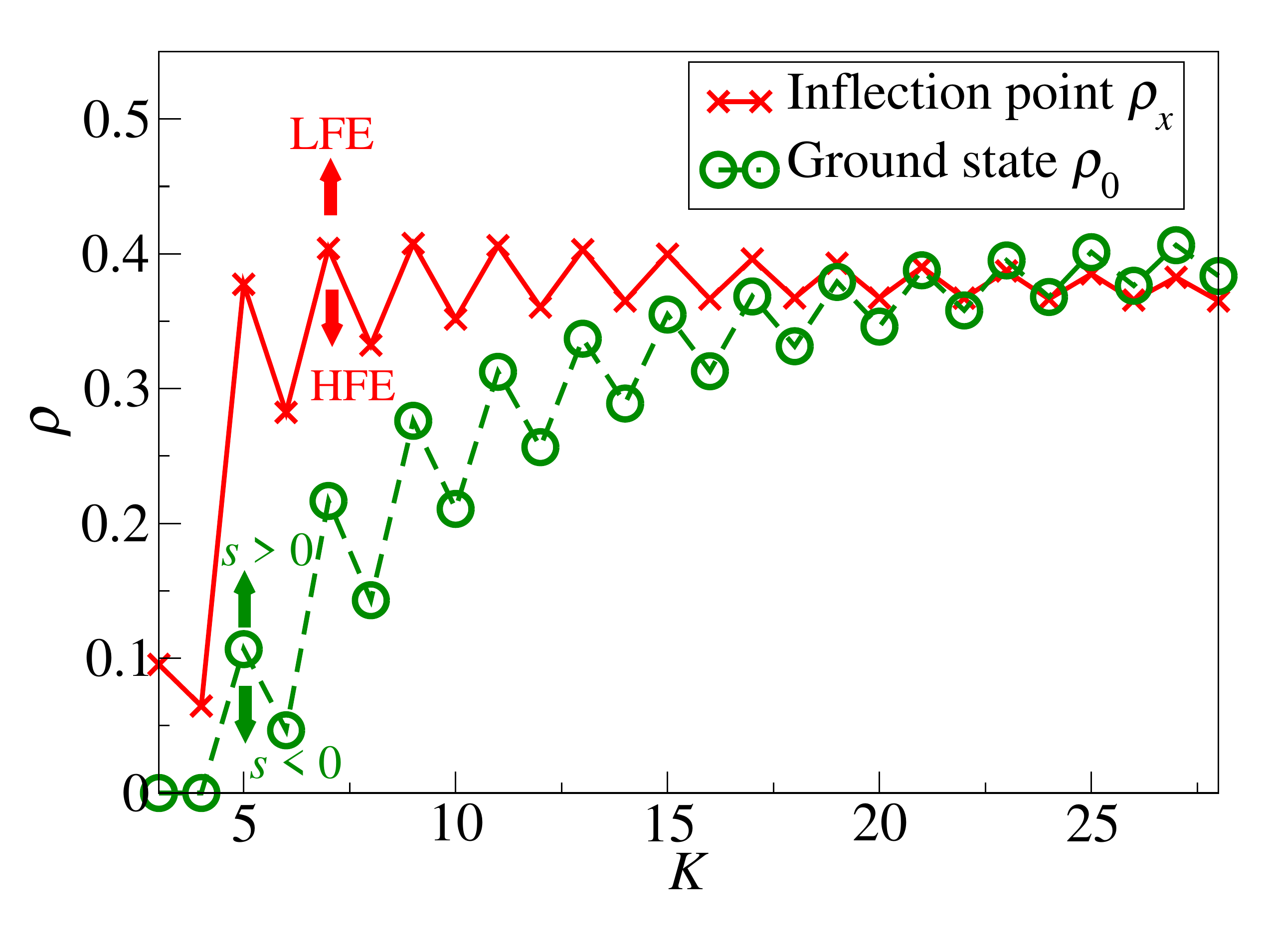}
  \caption{
    The energy density $\rho_x$ of the inflection point (crosses) and the minimum energy density $\rho_o$ (circles), for the RR graph ensemble of degree $K\in \{3, 4, \ldots, 28\}$. $\rho_x$ separates the lower-free-energy (LFE) branch from the higher-free-energy (HFE) branch; $\rho_o$ is determined by the condition of zero entropy density.
  }
      \label{fig:RhoValues}
\end{figure}

Coming back to the cases of negative entropy, we note that for cases of $K\ge 5$, there is a negative-entropy regime just before $\rho=0$, implying that states with $\rho\gtrsim 0$ have negative entropy and unphysical and thus are not the ground states of the system. This interpretation of the negative-entropy states with $\rho\gtrsim 0$ is consistent with our analysis of the unlikely presence of clique of size $n=4$ in RR graphs with $K=5$, shown in Sec.~\ref{sec:simpleprob}. This further implies that for cases of $K\ge 5$, the SDA ground states are those states with the minimal values of $\rho$ just beyond the negative entropy regime.
By allowing the possibility of replica-symmetry-breaking at the vicinity of $\rho_o$, the precise values of the minimum energy density $\rho_o$ may be further improved (see, e.g.,~\cite{kabashima1999statistical,Mezard2009information}). We leave this RSB exploration for future studies.

\begin{figure}
  \centering
  \includegraphics[angle=270, width=0.4\textwidth]{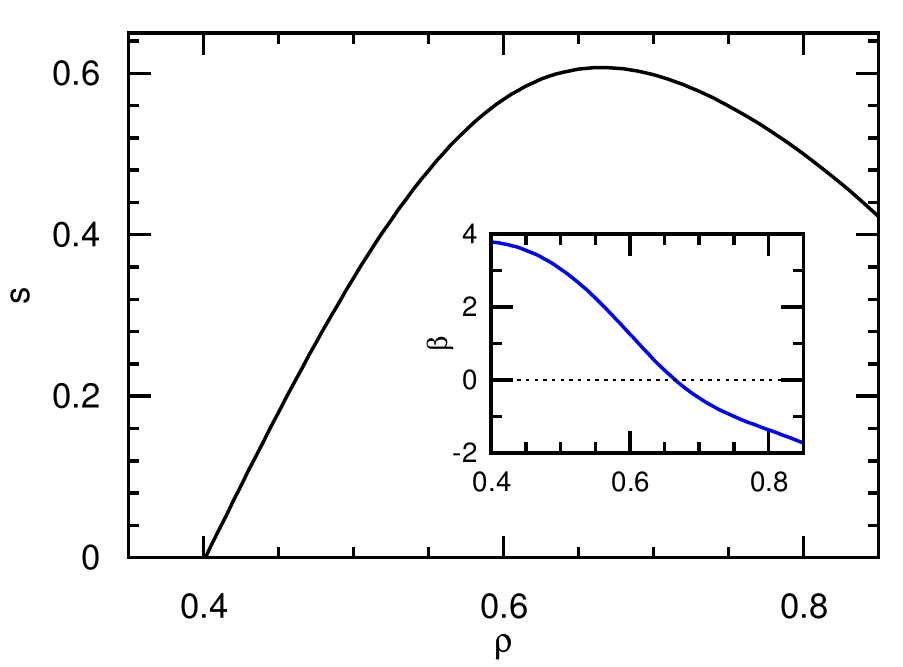}
  \caption{
    \label{fig:k25}
    The concave entropy density $s(\rho)$ for the RR graph ensemble of degree $K=25$. The inset shows the slope ${\rm d}s(\rho)/{\rm d}\rho$ of the entropy density.
  }
\end{figure}

We have determined the values of $\rho_x$ and $\rho_o$ for the RR graph ensembles with $K$ ranging from $K=3$ to $K=28$, see Fig.~\ref{fig:RhoValues}. Both $\rho_x$ and $\rho_o$ depend on the even-odd parity of $K$ and show oscillating behavior. We find that $\rho_x > \rho_o$ only for $K\leq 22$. When $K\geq 23$, we have $\rho_x < \rho_o$, so the entropy density is concave in the whole physically relevant range of $\rho\geq \rho_o$ (see Fig.~\ref{fig:k25} for the example of $K=25$).

\section{Solving the belief-propagation equation}
\label{sec:bpsolve}

In addition to the method used in the preceding section, we also employ conventional methods~\cite{Mezard2009information} for solving the BP equation [Eq.~(3) of the main text].

\subsection{With fixed inverse temperature $\beta$}

At a given fixed value of $\beta$, we iterate the BP equation on a single graph $G$ to obtain a fixed-point solution. At each elemental iteration process a vertex $j$ is randomly chosen from all $N$ vertices of the graph, and the cavity probability distributions $q_{j\rightarrow i}^{c_j, c_i}$ on the edges between $j$ and all its nearest neighbors $i$ are updated according to Eq.~(3). When $G$ is a RR graph we experience that this BP evolution converges to a fixed point within about $100 N$ elemental updates, and this fixed point is uniform in that the cavity probability distributions are identical for all the graph edges.
 
To get ensemble-averaged results for random graphs characterized by certain vertex degree profile, we also perform population dynamics simulations based on Eq.~(3). In the case of the RR graph ensemble, we first construct a long array of cavity probability distributions $q_{j\rightarrow i}^{c_j, c_i}$; then we repeatedly update it by (1) drawing $K\!-\!1$ cavity distributions uniformly at random from this array as inputs to Eq.~(3) to generate a new cavity distribution, and (2) replace an old cavity distribution in the array (chosen uniformly at random) by this new cavity distribution. This population dynamics also drives the population of cavity probability distributions to the uniform population (all the elements being identical) for the RR graph ensemble. The ensemble-averaged and single-graph BP results therefore are in complete agreement.

\subsection{With fixed energy density $\rho$}
\label{sec:C2}

To perform BP iteration at fixed energy density $\rho$, we need to slightly modify Eq.~(3) as follows
\begin{subequations}
  \label{eq:BPweight}
  \begin{align}
    w_{j\rightarrow i}^{0,0} & \equiv  w_{j\rightarrow i}^{0,1} = 
    \prod_{k\in \partial j\backslash i} 
    ( q_{k\rightarrow j}^{0,0}+q_{k\rightarrow  j}^{1,0}) \; ,
    \\
    w_{j\rightarrow i}^{1,0} & = \sum_{\bm{c}_{\partial j\backslash i}}
    \Theta\bigl(\sum\limits_{k\in\partial j\backslash i} c_k 
    -\frac{d_j}{2}\bigr) \prod\limits_{k\in \partial j\backslash i}
    q_{k\rightarrow j}^{c_k,1} \; ,
    \\
    w_{j\rightarrow i}^{1,1} & = \sum\limits_{\bm{c}_{\partial j\backslash i}} 
    \Theta\bigl(\sum\limits_{k\in\partial j\backslash i} c_k + 1 
    -\frac{d_j}{2} \bigr) \prod\limits_{k\in \partial j\backslash i} 
    q_{k\rightarrow j}^{c_k,1} \; ,
  \end{align}
\end{subequations}
where $w_{j\rightarrow i}^{0, 0}$, $w_{j\rightarrow i}^{0, 1}$, $w_{j\rightarrow i}^{1, 0}$, and $w_{j\rightarrow i}^{1, 1}$ are four auxiliary weight messages from vertex $j$ to its nearest neighbor $i$. We denote these four real quantities collectively as $\bm{w}_{j\rightarrow i}$. Similarly, we define the marginal weights $\bm{w}_j\equiv (w_j^{0}, w_{j}^1)$ of vertex $j$ as
\begin{subequations}
  \label{eq:weight}
  \begin{align}
    w_{j}^{0} & \equiv \prod_{k\in \partial j} 
    ( q_{k\rightarrow j}^{0,0}+q_{k\rightarrow  j}^{1,0}) \; ,
    \\
    w_{j}^{1} & \equiv \sum\limits_{\bm{c}_{\partial j}} 
    \Theta\bigl(\sum\limits_{k\in\partial j} c_k  
    -\frac{d_j}{2} \bigr) \prod\limits_{k\in \partial j} 
    q_{k\rightarrow j}^{c_k,1} \; ,
  \end{align}
\end{subequations}

In each BP iteration the following actions are taken: (1) we update the output messages $\bm{w}_{j\rightarrow i}$ and $\bm{w}_{i\rightarrow j}$ for each pair of edges $(i, j)$ of the graph according to Eq.~(\ref{eq:BPweight}),  and the marginal weights $\bm{w}_j$ for all the vertices $j$ according to Eq.~(\ref{eq:weight}); (2) and determine the value of the inverse temperature $\beta$ as the root of the following equation
\begin{equation}
  \label{eq:beta}
  \rho = \sum\limits_{j=1}^{N} \frac{e^{-\beta} w_j^1}{e^{-\beta} w_j^1 + w_j^0}
  \; ;
\end{equation}
and (3) we re-calculate the cavity probability distributions $q_{j\rightarrow i}$ between all the nearest-neighboring vertices using the new $\beta$:
\begin{subequations}
  \label{eq:qibeta}
  \begin{align}
    q_{j\rightarrow i}^{0,0} & \equiv  q_{j\rightarrow i}^{0,1} 
    = \frac{1}{z_{j\rightarrow i}} w_{j\rightarrow i}^{0, 0}\; , \\
    q_{j\rightarrow i}^{1,0} & =
    \frac{e^{-\beta}}{z_{j\rightarrow i}} w_{j\rightarrow i}^{1, 0} \; , \\
    q_{j\rightarrow i}^{1,1} & =
    \frac{e^{-\beta}}{z_{j\rightarrow i}} w_{j\rightarrow i}^{1, 1} \; ,
  \end{align}
\end{subequations}
where $z_{j\rightarrow i}$ is the normalization constant.

Similar to the discussions in the preceding subsection, we also iterate the modified BP equations (\ref{eq:BPweight})--(\ref{eq:qibeta}) by population dynamics to get ensemble-averaged results for the random SDA problem. For the RR graph ensembles the population dynamics results are in full agreement with BP results on single graph instances.

\section{Some simple probabilistic arguments concerning entropy and energy}
\label{sec:simpleprob}

The entropy density of the SDA problem is revealed by the cavity theory to be non-concave. Here we present a simple probabilistic theory to further confirm this non-concavity.

Consider a random regular graph of degree $K$. The total number of occupation configurations $\bm{c}$ with $N \rho$ occupied vertices and $(1-\rho) N$ empty vertices is simply $C_{N}^{N \rho}$. The probability that a randomly chosen configuration from this subset being an alliance is
\begin{equation}
\biggl[
  \sum\limits_{d\geq \frac{K}{2}}^{K} \frac{K!}{d! (K-d)!} 
  \rho^{d} (1-\rho)^{K-d}
  \biggr]^{N \rho} \; .
\end{equation}
Therefore the mean number of alliances with a given relative size $\rho$ is estimated to be 
\begin{equation}
  \Omega(\rho) = C_{N}^{N \rho}
  \biggl[
    \sum\limits_{d\geq K/2}^{K} \frac{K!}{d! (K-d)!} \rho^{d} (1-\rho)^{K-d}
    \biggr]^{N \rho} \; .
\end{equation}
At the thermodynamic limit $N\rightarrow \infty$, the entropy density $s(\rho) \equiv \frac{1}{N} \ln \Omega(\rho)$ is then estimated to be
\begin{eqnarray}
  s(\rho)  & = & -\rho \ln \rho - (1-\rho) \ln (1-\rho) 
  \label{eq:sKtheory}
  \\
  & &  + \rho \ln \biggl[ \sum\limits_{d\geq \frac{K}{2}}^{K} \frac{K!}{d! (K-d)!} 
    \rho^{d} (1-\rho)^{K-d}\biggr] \; .  \nonumber
 \end{eqnarray}

\begin{figure}
  \centering
  \includegraphics[angle=270,width=0.4\textwidth]{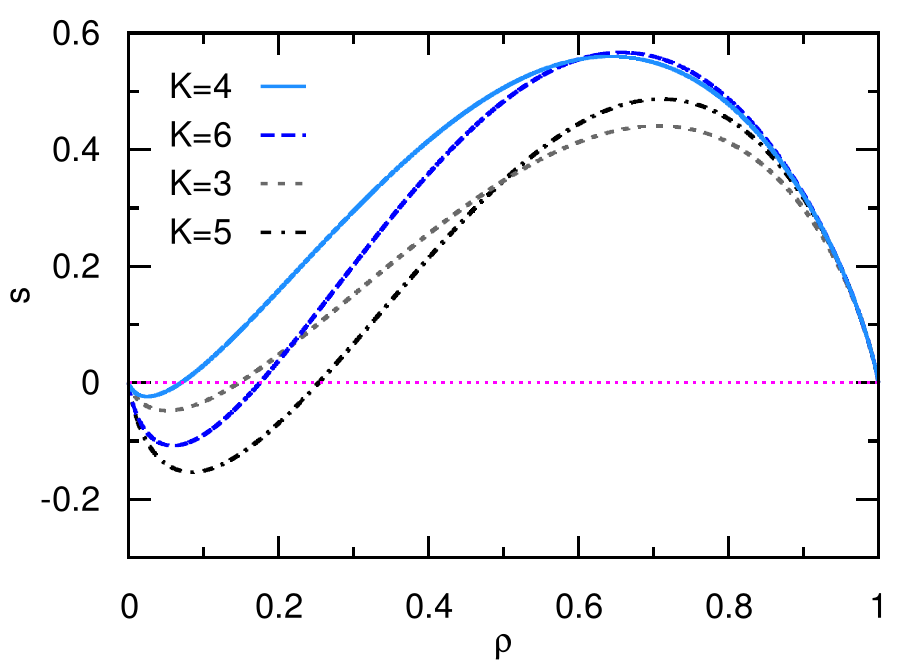}
  \caption{
    \label{fig:RRtheory}
    The non-concave entropy density function $s(\rho)$ as predicted by the simple probabilistic theory [Eq.~(\ref{eq:sKtheory})] for the RR graph ensemble of degree $K\in \{3, 4, 5, 6\}$.
  }
\end{figure}

This simple probabilistic theory predicts that the entropy density function $s(\rho)$ is convex when $\rho$ is small, see Fig.~\ref{fig:RRtheory}. Furthermore it predicts $s(\rho)$ to be negative for $0 < \rho < \rho_o$ with $\rho_o$ being some $K$-dependent threshold value, which means that there should not be any defensive alliance with relative size $\rho < \rho_o$. These predictions are in qualitative agreement with the results of the RS cavity theory.

The size $n$ of a minimum alliance for a $3$- or $4$-RR graph is found by the SA algorithm to be $n=3$, namely, the minimum alliance is a triangle. On the other hand, both theory and SA simulations suggest that the minimum alliance size of a $5$-RR graph is extensive. One would wonder why a small change in the value of $K$ from $4$ to $5$ results in an extensive gap in the size of alliance size $n$. Given a $5$-RR graph of large size $N$, why should we not expect to find a clique of size $n=4$ (with each vertex connecting to all the three other vertices of this clique) to serve as a minimum alliance? Here we offer an intuitive explanation. The expected number of a clique of size $n=4$ in a $5$-RR graph is
  \begin{equation}
  \label{eq:clique}
    \left(\begin{array}{c}N \\ 4\end{array}\right)
      \frac{\left(\begin{array}{c}N-4 \\ 2\end{array}\right)}{\left(\begin{array}{c}N-1 \\ 5\end{array}\right)}
      \frac{\left(\begin{array}{c}N-4 \\ 2\end{array}\right)}{\left(\begin{array}{c}N-2 \\ 4\end{array}\right)}
      \frac{\left(\begin{array}{c}N-4 \\ 2\end{array}\right)}{\left(\begin{array}{c}N-3 \\ 3\end{array}\right)}
      \frac{\left(\begin{array}{c}N-4 \\ 2\end{array}\right)}{\left(\begin{array}{c}N-4 \\ 2\end{array}\right)}
      \approx \frac{90}{N^2} \; ,
  \end{equation}
  which is vanishingly small as $N\rightarrow \infty$ and therefore will not be observed in a typical $5$-RR graph.
These results are consistent with the negative-entropy regime at $\rho\gtrsim 0$ for $K=5$, obtained by the mean-field cavity method (see Fig.~\ref{fig:entropyK3-10}), they also suggest the unlikely presence of cliques of size $n=4$.

Applying the same analysis of Eq.~(\ref{eq:clique}) to $3$-RR and $4$-RR graphs we find that the expected number of triangles is of order$O(1)$:
\begin{equation}
  \label{eq:clique3}
  \left(\begin{array}{c}N \\ 3\end{array}\right)
    \frac{\left(\begin{array}{c}N-3 \\ K-2\end{array}\right)}{\left(\begin{array}{c}N-1 \\ K\end{array}\right)}
    \frac{\left(\begin{array}{c}N-3 \\ K-2\end{array}\right)}{\left(\begin{array}{c}N-2 \\ K-1\end{array}\right)}
    \frac{\left(\begin{array}{c}N-3 \\ K-2\end{array}\right)}{\left(\begin{array}{c}N-3 \\ K-2\end{array}\right)}
    \approx \frac{K (K-1)^2}{6} \; , 
\end{equation}
where $K=3$ or $K=4$. Therefore triangles will be observed in these graphs. These results are again consistent with the profile of $s(\rho)$ for the cases of $K=3,4$, which is always characterized by positive entropy, including those states with $\rho\sim 1/N$, i.e. states with small loops.

\section{Stability of the Jacobian matrix of cavity probabilities}
\label{sec:stability}

To examine the stability of the recursion relation of cavity probabilities around the BP fixed-point solution with fixed $\beta$, we examine the stability of the equation with respect to small perturbations $\delta q^{(c_1, c_2)}$, by considering the largest absolute eigenvalue of the corresponding Jacobian matrix. We first differentiate Eq.~(3) of the main text  as follows:
\widetext
\begin{subequations}
  \label{eq_deltaBP}
  \begin{align}
    \delta q_{j\rightarrow i}^{0,0} & \equiv  \delta q_{j\rightarrow i}^{0,1} 
    = \frac{1}{z_{j\rightarrow i}} \sum_{l \in \partial j\backslash i}
    \bigl[\prod_{k\in \partial j\backslash i, l}
      ( q_{k\rightarrow j}^{0,0}+q_{k\rightarrow  j}^{1,0}) \bigr]
    (\delta q_{l\rightarrow j}^{0,0} + \delta q_{l\rightarrow  j}^{1,0}) 
      -\frac{1}{z^2_{j\rightarrow i}}\bigl[\prod_{k\in \partial j\backslash i} 
      ( q_{k\rightarrow j}^{0,0}+q_{k\rightarrow  j}^{1,0})\bigr]
    \delta z_{j\rightarrow i} \; ,
    \\
    \delta q_{j\rightarrow i}^{1,0} & 
    = \frac{e^{-\beta}}{z_{j\rightarrow i}}
    \sum_{l \in \partial j\backslash i} \Bigl[\sum_{\bm{c}_{\partial j\backslash i}}
      \Theta\bigl(\sum\limits_{k\in\partial j\backslash i} c_k -\frac{d_j}{2}\bigr) 
      \prod\limits_{k\in \partial j\backslash i, l} q_{k\rightarrow j}^{c_k,1} 
      \Bigr] \delta  q_{l\rightarrow j}^{c_l,1} 
    -\frac{1}{z^2_{j\rightarrow i}}
    \Bigl[\sum_{\bm{c}_{\partial j\backslash i}}
      \Theta\bigl(\sum\limits_{k\in\partial j\backslash i} c_k
      -\frac{d_j}{2}\bigr) \prod\limits_{k\in \partial j\backslash i}
      q_{k\rightarrow j}^{c_k,1}\Bigr]
    \delta z_{j\rightarrow i} \; ,
    \\
    \delta q_{j\rightarrow i}^{1,1} & 
    = \frac{e^{-\beta}}{z_{j\rightarrow i}}  \sum_{l \in \partial j\backslash i}
    \Bigl[\sum_{\bm{c}_{\partial j\backslash i}}
      \Theta\bigl(\sum\limits_{k\in\partial j\backslash i} c_k +1 
      -\frac{d_j}{2}\bigr) 
      \prod\limits_{k\in \partial j\backslash i, l} 
      q_{k\rightarrow j}^{c_k,1}\Bigr] \delta  q_{l\rightarrow j}^{c_l,1} 
    -\frac{1}{z^2_{j\rightarrow i}}
    \Bigl[\sum_{\bm{c}_{\partial j\backslash i}}
      \Theta\bigl(\sum\limits_{k\in\partial j\backslash i} c_k +1 
      -\frac{d_j}{2}\bigr) 
      \prod\limits_{k\in \partial j\backslash i} q_{k\rightarrow j}^{c_k,1}\Bigr]
    \delta z_{j\rightarrow i} \; ,
  \end{align}
\end{subequations}
where the change of normalization constant is 
\begin{align}
  \delta z_{j\rightarrow i} &=  \sum_{l \in \partial j\backslash i} 
  \Bigl\{ 2\bigl[\prod_{k\in \partial j\backslash i, l}
    ( q_{k\rightarrow j}^{0,0}+q_{k\rightarrow  j}^{1,0}) \bigr]
  (\delta q_{l\rightarrow j}^{0,0} + \delta q_{l\rightarrow  j}^{1,0}) 
  +  e^{-\beta}\bigl[\sum_{\bm{c}_{\partial j\backslash i}}
    \Theta\bigl(\sum\limits_{k\in\partial j\backslash i} c_k -\frac{d_j}{2}\bigr) 
    \prod\limits_{k\in \partial j\backslash i, l} q_{k\rightarrow j}^{c_k,1} \bigr]
  \delta  q_{l\rightarrow j}^{c_l,1}
  \nonumber\\
  & \quad 
  +
  e^{-\beta}\bigl[\sum_{\bm{c}_{\partial j\backslash i}}
    \Theta\bigl(\sum\limits_{k\in\partial j\backslash i} c_k +1 
    -\frac{d_j}{2}\bigr) 
    \prod\limits_{k\in \partial j\backslash i, l} q_{k\rightarrow j}^{c_k,1} \bigr]
  \delta  q_{l\rightarrow j}^{c_l,1}
  \Bigr\}  \; .
\end{align}

We then re-write Eq.~(\ref{eq_deltaBP}) in terms of $a, b, c$ and $d$ given by Eq.~(\ref{eq:abcd}) to simplify the subsequent derivation, and assuming the uniformity of vertices in RR graphs:
\begin{subequations}
  \label{eq_deltaabcd}
  \begin{align}
    \delta b_{j\rightarrow i} & = \delta d_{j\rightarrow i} = 
    \frac{1}{z} \sum_{l \in \partial j\backslash i}(D_{bc}\delta c_{l\rightarrow j} 
    + D_{bb}\delta b_{l\rightarrow j}) - \frac{b}{z}\delta z \; ,
    \\
    \delta c_{j\rightarrow i} & = \frac{e^{-\beta}}{z}
    \sum_{l \in \partial j\backslash i}(D_{ca}\delta a_{l\rightarrow j} +
    D_{cb}\delta b_{l\rightarrow j})  - \frac{c}{z}\delta z \; ,
    \\
    \delta a_{j\rightarrow i} & = \frac{e^{-\beta}}{z} 
    \sum_{l \in \partial j\backslash i}(D_{aa}\delta a_{l\rightarrow j} 
    + D_{ab}\delta b_{l\rightarrow j})  - \frac{a}{z}\delta z \; ,
  \end{align}
\end{subequations}
where the coefficients are
\begin{subequations}
  \begin{align}
    D_{aa} &= \sum_{r=\ceil{\frac{K}{2}-1}-1}^{K-2}C_{r}^{K-2}a^{r} b^{K-r-2} \; ,
    \\
    D_{ab} &= \sum_{r=\ceil{\frac{K}{2}-1}}^{K-2}C_{r}^{K-2}a^{r} b^{K-r-2} \; ,
    \\
    D_{bb} &= D_{bc} = (c+d)^{K-2} \; ,
    \\
    D_{ca} &= \sum_{r=\ceil{\frac{K}{2}}-1}^{K-2}C_{r}^{K-2}a^{r} b^{K-r-2} \; ,
    \\
    D_{cb} &= \sum_{r=\ceil{\frac{K}{2}}}^{K-2}C_{r}^{K-2}a^{r} b^{K-r-2} \; .
  \end{align}
\end{subequations}
Since $a+b+c+d = 1$ and $b=d$, we have $\delta a + \delta b + \delta c + \delta d = \delta a + 2\delta b + \delta c = 0$, and therefore we can write all the equations in terms of $\delta a$ and $\delta b$ only. We first re-write Eq. (\ref{eq_deltaabcd}a) as
\begin{align}
  \delta b_{j\rightarrow i} & = 
  \delta d_{j\rightarrow i} = \frac{1}{z}
  \sum_{l \in \partial j\backslash i}D_{bb}(\delta b_{l\rightarrow j}
  + \delta c_{l\rightarrow j}) - \frac{b}{z}\delta z 
  \nonumber \\
  & =
  \frac{1}{z} \sum_{l \in \partial j\backslash i}D_{bb}(-\delta a_{l\rightarrow j} 
  -\delta b_{l\rightarrow j})  - \frac{b}{z}\delta z \; .
\end{align}
The variable $\delta z$ is then given by the following equation in terms of $\delta a$ and $\delta b$:
\begin{equation}
  \delta z  =  \sum_{l \in \partial j\backslash i}\Bigl[
    (e^{-\beta}D_{aa} -2D_{bb} + e^{-\beta}D_{ca})\delta a_{l\rightarrow j}
     + (e^{-\beta}D_{ab} -2D_{bb} + e^{-\beta}D_{cb})\delta b_{l\rightarrow j}
    \Bigr] \; . 
\end{equation}
Finally, we write down a $2\times 2$ Jacobian matrix as
\begin{align}
  \mathcal{M} =
  \frac{1}{z}
  \begin{pmatrix}
    e^{-\beta}D_{aa}-a(e^{-\beta}D_{aa} -2D_{bb} + e^{-\beta}D_{ca})
    \ \ \ &  \ \ \
    e^{-\beta}D_{ab}-a(e^{-\beta}D_{ab} -2D_{bb} + e^{-\beta}D_{cb})
    \\
    -D_{bb}-b(e^{-\beta}D_{aa} -2D_{bb} + e^{-\beta}D_{ca})
    \ \ \ & \ \ \
    -D_{bb}-b(e^{-\beta}D_{ab} -2D_{bb} + e^{-\beta}D_{cb})
  \end{pmatrix}
\end{align}
\endwidetext
such that
\begin{align}
  \begin{pmatrix}
    \delta a_{j\rightarrow i}
    \\
    \delta b_{j\rightarrow i}
  \end{pmatrix}
  =
  \sum_{l \in \partial j\backslash i}\mathcal{M} 
  \begin{pmatrix}
    \delta a_{l\rightarrow j}
    \\
    \delta b_{l\rightarrow j}
  \end{pmatrix}
  =
  (K-1) \mathcal{M}
  \begin{pmatrix}
    \delta a_{l\rightarrow j}
    \\
    \delta b_{l\rightarrow j}
  \end{pmatrix}
  \; .
\end{align}

Following the arguments in~\cite{rivoire2004glass}, when the largest absolute eigenvalue $|\lambda_{max}|$ of the Jacobian matrix $\mathcal{M}$ satisfies
\begin{align}
  \label{eq_liquid}
  (K-1)|\lambda_{max}| > 1 \; ,
\end{align}
the differences $(\delta a, \delta b)$ in the cavity probabilities diverge, which indicates the instability of the so-called liquid solution (the so-called modulation instability~\cite{rivoire2004glass}).

\begin{figure}
  \centering
  \includegraphics[width=0.4\textwidth]{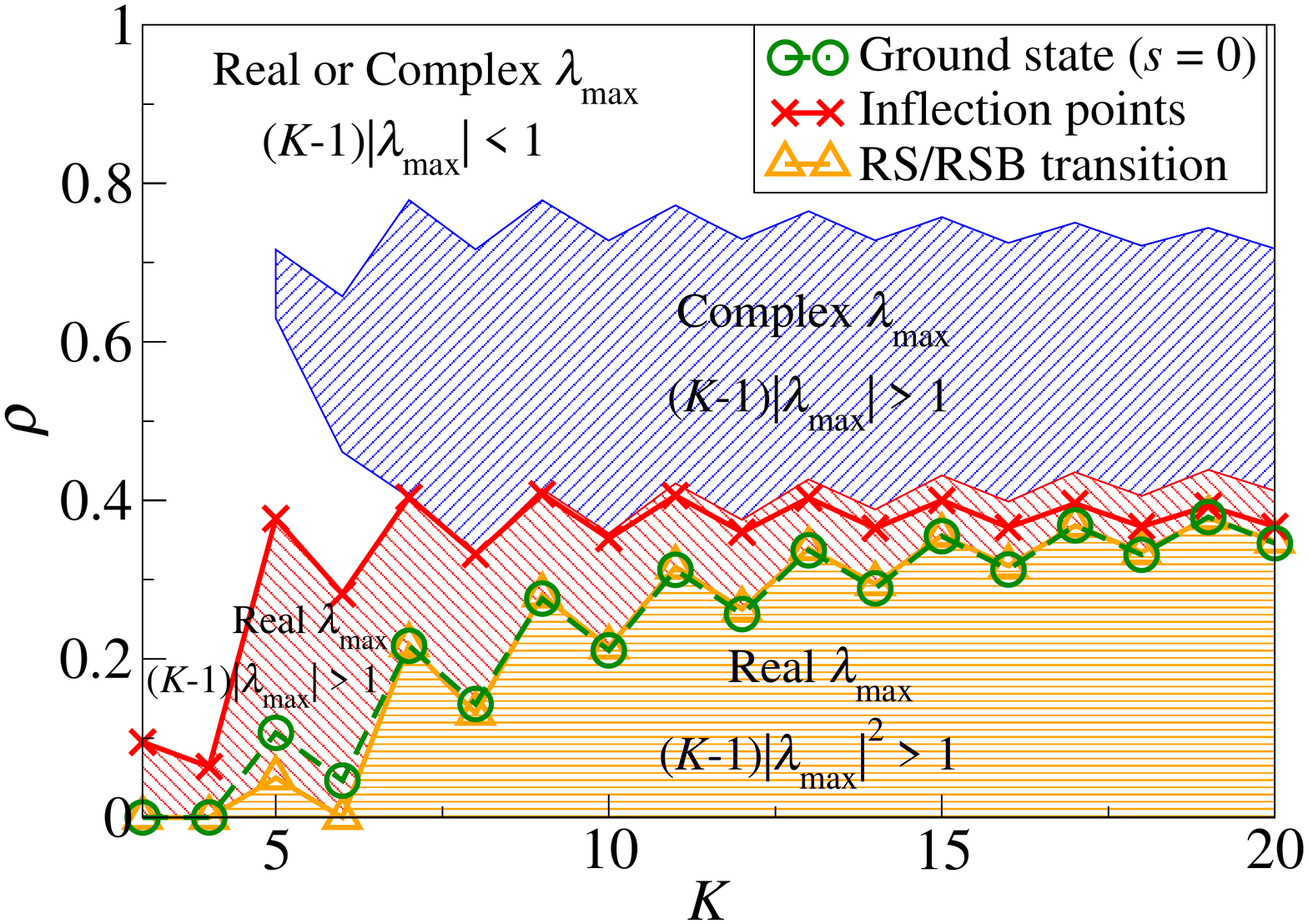}
  \caption{
    \label{fig:Kstability}
    Comparing the value of the energy density $\rho$ at the ground state ($\rho_o$ determined by entropy density $s=0$, circles), at the entropy inflection point ($\rho_x$, crosses), and at the RS/RSB (spin glass) transition point as determined by local stability analysis (triangles). The values of $\rho$ which satisfy Eq.~(\ref{eq_RSB}), i.e. the RSB phase, are marked in orange (horizontal stripes); those which satisfy Eq.~(\ref{eq_liquid}) with real eigenvalues $\lambda_{\rm max}$, i.e. the modulation phase, are marked in red (stripes with negative slopes); and values that satisfy Eq.~(\ref{eq_liquid}) with complex eigenvalues $\lambda_{\rm max}$ are marked in blue (stripes with positive slopes).
}
\end{figure}

On the other hand, when
\begin{align}
  \label{eq_RSB}
  (K-1)|\lambda_{max}|^2 > 1 \; ,
\end{align}
the variances $(\avg{(\delta a)^2}, \avg{(\delta b)^2})$ in the cavity probabilities diverge, which indicates the spin glass transition, i.e., the instability of a replica-symmetric (RS) solution to a replica-symmetry-breaking (RSB) solution.

As we can see in Fig.~\ref{fig:Kstability}, the values of $\rho$ of the RS/RSB spin-glass transition are consistent with (or just slightly above) the values of $\rho_o$ at the ground state, except for $K=5$ and $6$. These results are obtained without computing the entropy of the system. They imply that the higher-free-energy branch ($\rho_o < \rho < \rho_x$) of the RS cavity theory is locally stable. In other words, the predicted discontinuous phase transition identified in the main text between the high-energy phase and the ground-state phase is not due to the emergence of the RSB behavior but an effect associated with entropy inflection.

In addition, the values of $\rho$ with a real $\lambda_{\rm max}$ satisfying Eq.~(\ref{eq_liquid}) (i.e. the red region) are generally found below the inflection points; specifically, these $\rho$ values are consistent with the inflection points for $K=3,\ldots,6$. This implies that the inflection points roughly mark the onset of modulation instability, which may correspond to the fragmentation of the large alliances into smaller ones. This region is characterized by completely different SDA solutions, possibly with non-overlapping alliance members. On the other hand, we note that there is a large range of $\rho$ above the inflection points where the eigenvalues $\lambda_{\rm max}$ are complex and $|\lambda_{\rm max}|>1$ (i.e. the blue region). Nevertheless, since the eigenvalues are complex the instability on $(\delta a, \delta b)$ is rotational in nature, and neither cavity states $a$ nor $b$ vanish eventually. With an appropriate initial condition and a sufficiently slow adaptive iterative procedure, the iteration of the cavity equations lead to a uniform solution similar to the one found in the regime with $|\lambda_{\rm max}|<1$. In this region, different SDA solutions  with overlapping alliance members possibly  co-exist, leading to uniform cavity probabilities $a$ and $b$ on individual nodes.

If the energy density $\rho$ is kept fixed during the BP iterations, instead of the inverse temperature $\beta$,  we find that the modulation instability disappears, and only the spin glass RS/RSB instability remains (at the $\beta$/$\rho$ values identified before). For example, for the RR graph ensemble of degree $K=12$, the $\rho$-fixed RS population dynamics simulations always converge to the uniform BP fixed-point determined by Eqs.~(\ref{eq:BPsimple}a)-(\ref{eq:BPsimple}c), irrespective of the initial conditions, as long as $\rho \geq  0.263$. This stability threshold fully agrees with the theoretical prediction of the RS/RSB transition occurring at $\rho\approx 0.263$, which is only slightly above the predicted minimum energy density $\rho_o\approx 0.257$.

\section{The potential for a clustering transition}
\label{sec:cluster}

Besides the local stability analysis of Sec.~\ref{sec:stability}, we also check the possibility of a spin glass dynamical (clustering) transition in the SDA problem. We follow the theoretical method of~\cite{Mezard-Montanari-2006,Krzakala-etal-PNAS-2007} in this analysis. The corresponding first-step replica-symmetry-breaking (1RSB) results obtained by population dynamics simulations following Refs.~\cite{Mezard-Montanari-2006,Krzakala-etal-PNAS-2007} reveal that the complexity of the system is identical to zero for $\rho > \rho_x$ (with $\rho_x$ being the energy density of the inflection point), re-confirming that the discontinuous phase transition at $\rho>\rho_x$ as predicted by the mean field theory in the main text is not a spin glass transition but a phase transition resulting from the sigmoidal shape of the entropy function.
 
 Here we list the most essential message-passing equations used in the 1RSB population dynamics. A systematic review of the 1RSB theory can be found in~\cite{Mezard2009information}.
 
 To investigate the possibility of an ergodicity-breaking transition at $\rho> \rho_x$, we consider the 1RSB mean field theory at $y=\beta$, where $y$ is the inverse temperature at the level of macroscopic states. The distribution of the cavity probability function $q_{i\rightarrow j}$ among all macroscopic states is denoted as $Q_{i\rightarrow j}[q_{i\rightarrow j}]$. Let us first introduce an auxiliary probability functional
 \begin{equation}
   Q_{i\rightarrow j}^{c_i, c_j}[q_{i\rightarrow j}|\bar{q}_{i\rightarrow j}] \equiv
   \frac{Q_{i\rightarrow j}[q_{i\rightarrow j}]q_{i\rightarrow j}^{c_i,c_j}}{\bar{q}_{i\rightarrow j}^{c_i, c_j}} \; ,
 \end{equation}
 where the mean cavity probability is defined as $\bar{q}_{i\rightarrow j} \equiv \int \mathcal{D} q_{i\rightarrow j}\ Q_{i\rightarrow j}[q_{i\rightarrow j}] q_{i\rightarrow j}$ (averaging over all the possible cavity probability functions). At $y=\beta$ the mean cavity probabilities $\bar{q}_{i\rightarrow j}$ on all the edges of the graph satisfy the BP equation [see Eq.~(3) of the main text], and therefore they can be determined without knowing $Q_{i\rightarrow j}[q_{i\rightarrow j}]$. The functional $Q_{i\rightarrow j}^{c_i, c_j}[q_{i\rightarrow j}|\bar{q}_{i\rightarrow j}]$ can be understood as the conditional probability of drawing a cavity distribution $q_{i\rightarrow j}$ given the observed occupation states of vertex $i$ being $c_i$ and that of vertex $j$ being $c_j$ \cite{Mezard-Montanari-2006}. 

 At $y=\beta$ the self-consistent equation for this auxiliary probability functional is derived to be
 \widetext
 \begin{equation}
   \label{eq:1RSBq}
   Q_{i\rightarrow j}^{c_i,c_j}[q_{i\rightarrow j}| \bar{q}_{i\rightarrow j}] =
   \sum_{\bm{c}_{\partial i\backslash j}} 
   \Gamma_{i\rightarrow j}^{c_i,c_j}(\bm{c}_{\partial i\backslash j})
   \prod_{k\in\partial i\backslash j}\int\mathcal{D} q_{k\rightarrow i}
   Q_{k\rightarrow i}^{c_k,c_i}[q_{k\rightarrow i}|\bar{q}_{k\rightarrow i}] 
   \delta[q_{i\rightarrow j}-\hat{q}_{i\rightarrow j}]
   \; ,
 \end{equation}
 where 
 \begin{equation}
   \Gamma_{i\rightarrow j}^{c_i,c_j}(\bm{c}_{\partial i\backslash j})
   \equiv \frac{\delta^0_{c_i}\prod\limits_{k\in\partial i\backslash j}
     \bar{q}_{k\rightarrow i}^{c_k,0}+ \delta^1_{c_i}e^{-\beta}
     \Theta(\sum\limits_{k\in\partial i\backslash j}c_k+c_j-\frac{d_i}{2})
     \prod\limits_{k\in\partial i\backslash j}\bar{q}_{k\rightarrow i}^{c_k, 1}}{2
     \prod\limits_{k\in\partial i\backslash j}(\bar{q}_{k\rightarrow i}^{0,0}+
     \bar{q}_{k\rightarrow i}^{1,0})
     + e^{-\beta}\sum\limits_{c_j}\sum\limits_{\bm{c}_{\partial i\backslash j}}
     \Theta(\sum\limits_{k\in\partial i\backslash j} c_k+c_j-\frac{d_i}{2})
     \prod\limits_{k\in\partial i\backslash j}\bar{q}_{k\rightarrow i}^{c_k, 1}} \; ,
\end{equation}
and $\hat{q}_{i\rightarrow j}$ is a short-hand notation for the BP expression. The probability weights $\Gamma_{i\rightarrow j}^{c_i,c_j}(\bm{c}_{\partial i\backslash j})$ can be used to construct an occupation pattern $\bm{c}_{\partial i\backslash j}=\{c_k : k\in \partial i\backslash j\}$ for a focal vertex $i$, and then one can get a set of samples $q_{i\rightarrow j}$ following Eq.~(\ref{eq:1RSBq}) to represent $Q_{i\rightarrow j}^{c_i, c_j}[q_{i\rightarrow j} | \bar{q}_{i\rightarrow j}]$.

For the RR graph ensembles the 1RSB population dynamics simulations carried out for $\rho > \rho_x$ all evolve to the trivial fixed point of all the probability functionals $Q_{i\rightarrow j}^{c_i, c_j}[q_{i\rightarrow j} | \bar{q}_{i\rightarrow j}]$ and $Q_{i\rightarrow j}[q_{i\rightarrow j}]$ being Dirac's $\delta$-functionals. This indicates that the system has only a single equilibrium macroscopic state at energy density $\rho > \rho_x$.

The same 1RSB analysis, based on population dynamics, may be carried out for $\rho < \rho_x$ to determine the precise value of the spin glass dynamical transition point; this is beyond the scope of the current study and will be the subject of future research.

\section{The Clamp-Alliance (CA)  algorithm}
\label{sec:CA}

Here we present the pseudo-code of the CA algorithm. Algorithm~\ref{alg:ca} is based on the modified BP message-passing protocol (see Sec.~\ref{sec:C2}). The inverse temperature $\beta$ is adjusted by solving Eq.~(\ref{eq:beta}) after each BP iteration. The code of CA is accessible from the webpage {\tt power.itp.ac.cn/\~{}zhouhj/codes.html}.
 
The performance of the CA algorithm is not sensitive to the precise value of objective density $\rho$. The CA results reported in Table I of the main text were obtained by setting the objective relative size $\rho=\rho_o$, with $\rho_o$ being the estimated minimum energy density by the RS mean field theory.  If the value of $\rho_o$ is unknown, one can simply run the CA algorithm for a set of different objective $\rho$ values and choose the minimum-size alliance set $A$ obtained from these different trials.

For the alliance solutions $A$ obtained by the CA algorithm for RR graphs, the subgraph induced by the vertices of each of these alliances forms only a single connected component.

 \begin{algorithm}
   \caption{
     \label{alg:ca}
     Clamp-Alliance (CA) for the minimum strong defensive alliance problem. The output of CA is a vertex set (alliance) $A$ such that each vertex $i\in A$ has at least one half of its nearest-neighbor vertices in $A$.
   }
   \begin{algorithmic}[100]
     \State \textbf{Input}. A connected graph $G$ of $N$ vertices $i \in \{1, 2, \ldots, N\}$ and $M$ edges $(i, j)$ between pairs of vertices $i$ and $j$.
     \State \textbf{Initialize}. Set $S=\{1, 2, \ldots, N\}$; randomly assign cavity probability distributions $q_{i\rightarrow j}^{c_i, c_j}$ and $q_{j\rightarrow i}^{c_j, c_i}$ for all the edges $(i, j)$; set decimation fraction $\eta$ (e.g., $\eta=0.005$); set objective relative size $\rho$ of alliance;  set iteration number $t$ (e.g., $t=10$) of belief-propagation.
     \While{$S\neq \emptyset$}
     \Comment{reduce alliance size by BP-guided decimation}
     \begin{enumerate}
     \item[1.] Set $A=S$.
     \item[2.] Repeat $t$ times the iteration of the modified BP equation [see Eqs.~(\ref{eq:BPweight}) and (\ref{eq:qibeta})] on all the edges between the vertices of $S$, adjusting the inverse temperature $\beta$ of each BP iteration to satisfy condition (\ref{eq:beta}).
     \item[3.] Compute the occupation probabilities $q_i$ for all the vertices $i\in S$ according to $q_i=\frac{e^{-\beta} w_i^1}{\bigl(e^{-\beta} w_i^1+w_i^0\bigr)}$ [see Eq.~(\ref{eq:weight})], and then rank these vertices in increasing order of $q_i$.
     \item[4.] Delete the top fraction $\eta$ of the vertices $i\in S$ (which have the smallest $q_i$ values) from $S$.
     \item[5.] Repeatedly delete a vertex $j\in S$ from $S$ if $j$ has less than $d_j/2$ nearest neighbors in $S$, until no more vertices need to be deleted.
     \end{enumerate}
     \EndWhile
     \For{every vertex $i \in A$ in a random order}
     \Comment{refine the alliance}
     \begin{enumerate}
     \item[1.] Set $S=A$.
     \item[2.] Delete $i$ from $S$.
     \item[3.] Repeatedly delete a vertex $j\in S$ from $S$ if $j$ has less than $d_j/2$ nearest neighbors in $S$, until no more vertices need to be deleted.
     \item[4.] If $S\neq \emptyset$, then set $A=S$.
     \end{enumerate}
     \EndFor
   \end{algorithmic}
 \end{algorithm}

\end{document}